\documentstyle[amssymb,epsfig,epic,eepic]{myfbs}

\newcommand{\be}{\begin{equation}}
\newcommand{\ee}{\end{equation}}
\newcommand{\bea}{\begin{eqnarray}}
\newcommand{\eea}{\end{eqnarray}}
\newcommand{\mel}[3]{<\!#1\,|\,#2\,|\,#3\!>}

\newcommand{\scp}[2]{\mbox{$<\!#1\,|\,#2\!>$}}
\newcommand{\bra}[1]{\mbox{$<\!#1\,|$}}
\newcommand{\ket}[1]{\mbox{$|\,#1\!>$}}
\newcommand{\sep}[2]{\mbox{$|\, #1\!><\! #2 \, |$}}

\newcommand{\itrm}{\rm}

\title{Bayesian Inverse Quantum Theory\thanks{Dedicated to 
Professor Achim Weiguny
on the occasion of his retirement}}
\author{J\"org C. Lemm\thanks{lemm@uni-muenster.de}, Joerg Uhlig}
\institute{Institut f\"ur Theoretische Physik I,
Wilhelm--Klemm--Str.9,
D--48149 M\"unster, Germany}

\runningauthor{J.C.\ Lemm, J.\ Uhlig}
\runningtitle{Bayesian Inverse Quantum Theory}
\SetFinalPage{26}

\begin{document}

\maketitle
\begin{abstract}
A Bayesian approach is developed to determine
quantum mechanical potentials from empirical data. 
Bayesian methods, combining empirical measurements
and {\it a priori} information,
provide flexible tools for such empirical learning problems.
The paper presents the basic theory, 
concentrating in particular on measurements of particle coordinates
in quantum mechanical systems at finite temperature.
The computational feasibility of the approach
is demonstrated by numerical case studies.
Finally, it is shown how the approach can be generalized 
to such many--body and few--body systems
for which a mean field description is appropriate.
This is done by means of 
a Bayesian inverse Hartree--Fock approximation.
\end{abstract}

\tableofcontents

\section{Introduction}

The problem addressed in this paper
is the reconstruction of Hamiltonians of quantum systems
from observational data.
Finding such ``causes''  or ``laws''  
from a finite number of observations 
constitutes an {\it inverse problem} and 
is typically {\it ill--posed} in the sense of Hadamard 
[1--8].

Two research fields deal in particular
with the reconstruction of potentials from spectral data
(energy measurements):
{\it inverse spectral theory}
and
{\it inverse scattering theory}.
Inverse spectral theory 
characterizes the kind of 
data necessary, in addition to a given spectrum,
to determine the potential
[7, 9--13]. 
(See also Sect.\ \ref{need-for-prior}.)
Inverse scattering theory, in particular,
considers, in addition to the spectrum, boundary data
obtained `far away' from the scatterer.
Those can be, for example, phase shifts obtained from scattering experiments
\cite{Chadan-Colton-Paivarinta-Rundell-1997,Newton-1989,Chadan-Sabatier-1989}.

In this paper, contrasting those two approaches, we will 
not exclusively be interested in spectral data,
but will develop a formalism which allows 
to extract information from quite heterogeneous empirical data.
In particular, we will consider in more detail
the situation where the position of a quantum mechanical particle
has been measured a finite number of times.

Due to increasing computational resources,
the last decade has also seen
a rapidly growing interest in
applied empirical learning problems.
They appear as
density estimation, regression or classification
problems
and include, just to name a few,
image reconstruction,
speech recognition, 
time series prediction,
and object recognition. 
Many disciplines,
like applied statistics,
artificial intelligence,
computational and statistical learning theory,
statistical physics, 
and also psychology and biology,
contributed in developing
a variety of learning algorithms,
including for example
smoothing splines \cite{Wahba-1990},
regularization and kernel approaches \cite{Vapnik-1982},
support vector machines \cite{Vapnik-1995,Vapnik-1998},
generalized additive models \cite{Hastie-Tibshirani-1990},
projection pursuit regression \cite{Huber-1985}, 
expert systems and decision trees \cite{Breiman-Friedman-Olshen-Stone-1993},
neural networks \cite{Bishop-1995b},
and graphical models \cite{Lauritzen-1996}.

Recently, their has been much work devoted
to the comparison and unification of methods
arising from different disciplines.
(For an overview and comparison of methods
see for example \cite{StatLog-1994}.)
Hereby, especially the {\it Bayesian approach} to statistics
proved to be useful as a unifying framework for empirical learning
[22, 25--36].
Bayesian approaches put special emphasis on 
{\it a priori} information
which always has to accompany empirical data
to allow successful learning.

The present paper is written from a Bayesian perspective.
In particular,
{\it a priori} information will be implemented 
in form of stochastic processes \cite{Doob-1953}.
Compared to parametric techniques
this has the advantage, 
that {\it a priori} information 
can typically be controlled more explicitly.
Technically, this approach is intimately related
to the well known Tikhonov regularization 
\cite{Tikhonov-1963,Tikhonov-Arsenin-1977}.
For an outline of the basic principles see also \cite{Lemm-1999b}.

The paper is organized as follows:
Sect.\ \ref{bayesian} gives a short introduction to Bayesian statistics.
Sect.\ \ref{InvQS} applies the Bayesian approach to 
quantum mechanics and quantum statistics,
with Sect.\ \ref{likelihood-model} concentrating
on the treatment of empirical data for quantum systems
and Sect.\ \ref{prior-models} 
discussing the implementation of {\it a priori} information.
Sect.\ \ref{numerics} presents two numerical case studies,
the first dealing with the approximation of approximately periodic potentials,
the second with inverse two--body problems.
Sect.\ \ref{Inv-many-body} shows how
the approach can be applied to many--body systems,
including the fundamentals of an inverse version of Hartree--Fock theory.
Finally, Sect.\ \ref{conclusions} concludes the paper.

\section{The Bayesian approach}
\label{bayesian}

\subsection{Basic notations}

A Bayesian approach is based upon two main ingredients: 
\begin{itemize}
\item[1.]
A model of Nature, i.e., a space ${\cal H}$ of hypotheses $h$ 
defined by their {\it likelihood} functions
$p(x|c,h)$.
Likelihood functions specify
the probability density
for producing outcome $x$ (measured value or dependent visible variable, 
assumed to be directly observable) 
under hypothesis $h$
(possible state of Nature or hidden variable, assumed to be not directly observable)
and condition $c$ 
(measurement device parameters or independent visible variable, 
assumed to be directly observable).

\item[2.] A {\it prior density} $p_0(h)$ = $p(h|D_0)$ 
defined over the space ${\cal H}$ of hypotheses,
$D_0$ denoting collectively all available {\it a priori} information.
\end{itemize}

Now assume (new) {\it training data} 
$D_T$ = $(x_T,c_T)$ = $\{(x_i,c_i)|1\le i\le n\}$
become available,
consisting of pairs
of measured values $x_i$ under known conditions $c_i$
(and unknown $h$). 
Then
{\it Bayes' theorem}
\be
p(h|D)
=\frac{p(x_T|c_T,h)\,p_0(h)}{p_0(x_T|c_T)}
\label{bayes-rule}
,
\ee
is used to update 
the prior density $p_0(h)$ = $p(h|D_0)$ to get the (new)
{\it posterior density} $p(h|D)$ = $p(h|D_T,D_0)$.
Here we wrote $D$ = $(D_T,D_0)$ to denote both,
training data and {\it a priori} information.
Assuming i.i.d.\ training data $D_T$ 
the likelihoods factorize
$p(x_T|c_T,h)$ = $\prod_i^n p(x_i|c_i,h)$.
Note that the denominator which appears in Eq.~(\ref{bayes-rule}),
i.e., $p_0(x_T|c_T)$ 
= $\int\! dh\, p(x_T|c_T,h)\,p_0(h)$,
is $h$--independent.
It plays the role of a normalization factor,
also known as evidence.
Thus, the terms in Eq.~(\ref{bayes-rule})
are named as follows, 
\be
{\rm posterior} 
= \frac{{\rm likelihood}\times{\rm prior}}{\rm evidence}
.
\ee

To make predictions,
a Bayesian approach aims at calculating
the {\it predictive density}
\be
p(x|c,D) 
= \int \!dh\, p(x|c,h)\, p(h|D)
,
\label{predictive}
\ee
which is a likelihood average
weighted by their posterior probability.
The $h$--integral can be extremely high dimensional,
and often, like in the case we are focusing on here, 
even be a functional integral \cite{Schulman-1981,Glimm-Jaffe-1987}
over a space of likelihood functions ${\cal H}$.
In as far as an analytical integration is not possible,
one has to treat the integral, for example, 
by Monte Carlo methods 
[30, 41--44] 
or in saddle point approximation
\cite{Berger-1980,Gelman-Carlin-Stern-Rubin-1995,de-Bruijn-1961,Bleistein-Handelsman-1986}.
Assuming the likelihood term $p(x|c,h)$ to be
slowly varying at the stationary point
the latter is also known
as {\it maximum posterior approximation}.
In this approximation the $h$--integration is effectively replaced 
by a maximization of the posterior, 
meaning the predictive density is approximated by 
\be
p(x|c,D) \approx p(x|c,h^*),
\ee
where
\be
h^*
= {\rm argmax}_{h\in{\cal H}} p(h|D)
= {\rm argmin}_{h\in{\cal H}} [-\ln p(h|D)]
.
\label{map-def}
\ee
The term $-\ln p(h|D)$
is also often referred to as 
(regularized) {\it error functional}
and indeed, a maximum posterior approximation
is technically equivalent to minimizing 
error functionals with Tikhonov regularization
[2--4, 47]. 
The difference between the Bayesian approach
and the classical Tikhonov regularization
is the interpretation of the extra term
as costs or as {\it a priori} information, respectively.

Within a
{\it maximum likelihood approach}
an optimal hypothesis $h$ is obtained by
maximizing only its training likelihood $p(x_T|c_T,h)$ 
instead of its complete posterior.
This is equivalent to a maximum posterior approximation 
with uniform prior density.
A maximum likelihood approach
can be used for hypotheses $h$ = $h(\xi)$,
parameterized by (vectors) $\xi$.
A maximum likelihood approach is possible
if that parameterization is restrictive enough
(and well enough adapted to the problem),
so no additional prior is required to 
allow generalization from training
to non--training data.
For completely flexible nonparametric approaches, however,
the prior term is necessary
to provide the necessary information
linking training data and (future) non--training data.
Indeed,
if every number $p(x|c,h)$ is considered
as a single degree of freedom
[restricted only by the positivity constraint $p(x|c,h) > 0$
and the normalization over $x$]
then, without {\it a priori} information,
training data contain no information
about non--training data.

\subsection{An Example: Regression}

Before applying the Bayesian framework to quantum theory, 
we shortly present one of its standard applications:
the case of (Gaussian) regression.
(For more details see for example \cite{Lemm-BFT-1999}.)
This also provides an example for
the relation between the Bayesian maximum posterior approximation
and the minimization of regularized error functionals.

\begin{figure}
\begin{center}
\epsfig{file=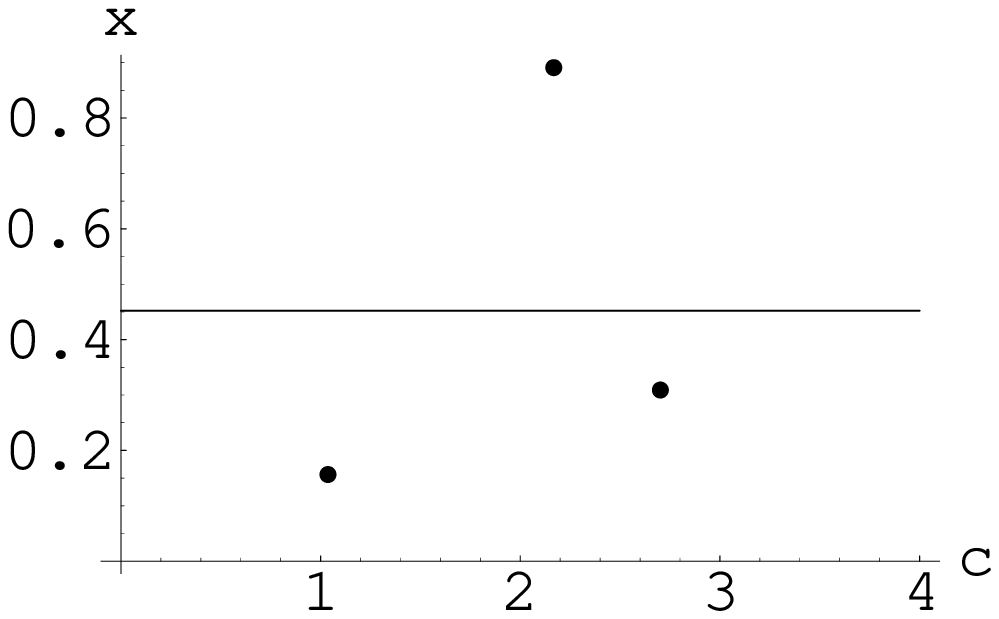, width= 40mm}
\epsfig{file=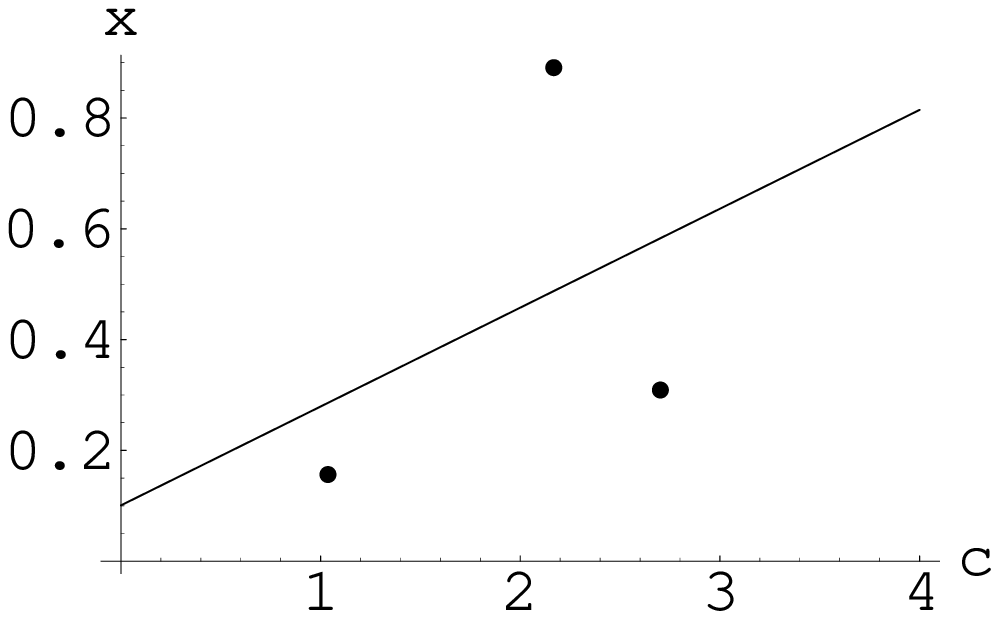, width= 40mm}
\epsfig{file=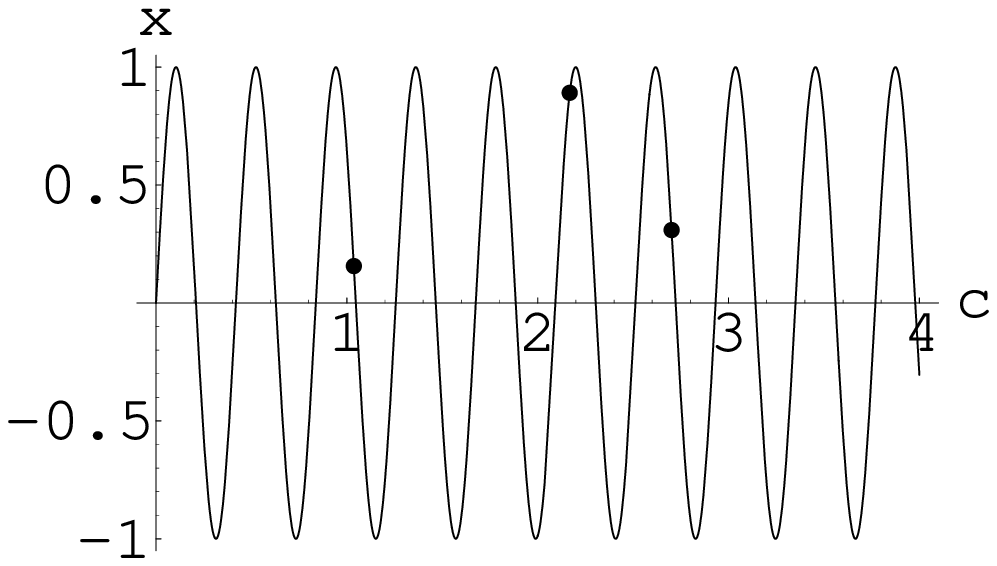, width= 40mm}
\end{center}
\caption{Examples of parametric regression
functions with increasing flexibility (with 3 data points).
L.h.s: A fitted constant $h(c)$ = $\xi$.
Middle: A linear $h(c)$ = $\xi_2 c +\xi_1$.
R.h.s: The function $h(c)$ = $\sin (\xi c)$ 
can fit an arbitrary number of data points 
(with different $c_i\ne 0$ and $|x_i|\le 1$)
well \cite{Vapnik-1995}.
Additional {\it a priori} information becomes especially important
for flexible approaches.
}
\label{sin}
\end{figure}

A regression model is a model with Gaussian likelihoods,
\be
p(x|c,h) = \frac{1}{\sqrt{2\pi}\sigma}e^{-\frac{(x-h(c))^2}{2\sigma^2}}
,
\ee
with fixed variance $\sigma^2$.
The function $h(c)$ is known as regression function.
(In regression, one often writes
$y$ for the dependent variable, which is $x$ in our notation, 
and $x$ for the ``condition'' $c$.)
Our aim is to determine an approximation for $h(c)$ 
using observational data $D$ = $\{(x_i,c_i)|1\le i\le n\}$.
Within a parametric approach
one searches for an optimal approximation 
in a space of parameterized regression functions
$h(c;\xi)$.
For example, in the simple cases of a constant or a linear regression 
such a parameterization would
be $h(c;\xi)$ = $\xi$ or $h(c;\xi_1,\xi_2)$ = $\xi_2 c + \xi_1$, respectively.
If the parameterization is restrictive enough
then a prior term $p_0(h)$ is not needed
and maximizing the likelihood over all data $D$
is thus equivalent to minimizing
the squared error,
\be
E_{\rm sq} (\xi) = \sum_i^n \Big( x_i-h(c_i;\xi) \Big)^2
.
\ee
There are, however,
also very flexible parametric approaches, which usually do require
additional {\it a priori} information.
An example of such a nonlinear one--parameter family 
has been given by Vapnik
and is shown in Fig. \ref{sin}.
Without additional {\it a priori} information,
which may for example restrict the number of oscillations,
such functions
can in most cases not be expected to lead to useful predictions.
Nonparametric approaches, which treat the
numbers $h(c)$ as single degrees of freedom,
are even more flexible and do always require a prior $p_0(h)$.
For nonparametric approaches such a prior
can be formulated in terms of the function values $h(c)$.
A technically very convenient choice is a Gaussian process prior
\cite{Williams-Rasmussen-1996,MacKay-1998},
\be
p_0(h) 
=
\left(\det \frac{\lambda}{2\pi}{\bf K}_0 \right)^\frac{1}{2}
e^{-\frac{\lambda}{2} 
\mel{h-h_0}{{\bf K}_0}{h-h_0}}
,
\label{gaussprior1}
\ee
with mean $h_0$, representing a 
reference or template for the regression function $h$,
and inverse covariance $\lambda{\bf K}_0$
given by a
real symmetric, positive (semi--)definite
operator 
scaled by $\lambda$ and acting on functions $h$.
The operator ${\bf K}_0$ defines the scalar product,
\be
\mel{h-h_0}{{\bf K}_0}{h-h_0}
=
\int \!dc \, dc^\prime \,
[h(c)-h_0(c)]
\, {\bf K}_0(c,c^\prime) \,
[h(c^\prime )-h_0(c^\prime)]
.
\ee
Typical priors enforce the regression function $h$ to be smooth.
Such smoothness priors are implemented by choosing differential operators
for ${\bf K}_0$.
For example, taking for ${\bf K}_0$ 
the negative Laplacian
and choosing a zero mean $h_0$ = $0$,
yields
\be
\mel{h-h_0}{{\bf K}_0}{h-h_0}
=
- \int \!dc \, 
h(c)
\, \frac{\partial^2 }{\partial c^2}
h(c)
=
\int \!dc \, 
\left(\frac{\partial h(c)}{\partial c}
\right)^2
,
\label{smooth1}
\ee
where we integrated by parts
assuming vanishing boundary terms.
In statistics one often prefers inverse covariance operators 
with higher derivatives to obtain smoother regression functions
[16, 50--55]. 
An example of such a prior with higher derivatives
is a ``Radial Basis Functions'' prior
with the pseudo-differential operator
${\bf K}_0$ = $\exp{(-{\sigma_{\rm RBF}^2}{\Laplace}/2)}$
as inverse covariance.

Maximizing the predictive density $p(x|c,D)$
for a Gaussian prior (\ref{gaussprior1})
is equivalent
to minimizing the regularized error functional
\be
E_{\rm reg} (h)
=
\frac{1}{2}\sum_i^n \Big( x_i-h(c_i) \Big)^2
+ \frac{\lambda^\prime}{2}\mel{h-h_0}{{\bf K}_0}{h-h_0}
.
\ee
The ``regularization'' parameter
$\lambda^\prime$ = $\lambda \sigma^2$,
representing a so called {\it hyperparameter},
controls the balance between empirical data
and {\it a priori} information.
In a Bayesian framework 
one can include a hyperprior $p(\lambda)$
and either integrate over $\lambda$
or determine an optimal $\lambda$ in maximum posterior approximation
\cite{Bishop-1995b,Berger-1980}.
Alternative ways to determine $\lambda$ are
crossvalidation techniques\cite{Wahba-1990},
the discrepancy and 
the self--consistent method \cite{Honerkamp-Weese-1990}.
For example in the case of a smoothness prior,
a larger $\lambda^\prime$ 
will result in a smoother regression function $h^*$.  
It is typical for the case of regression 
that the regularized error functional $E_{\rm reg} (h)$
is quadratic in $h$.
It is therefore easily minimized
by setting the functional derivative with respect to $h$
to zero, i.e.,
$\delta E_{\rm reg}/\delta h$ 
=
$\delta_h E_{\rm reg}$
= 0.
This stationarity equation is then linear in $h$ 
and thus has a unique solution $h^*$.
(This is equivalent to so called kernel methods
with kernel ${\bf K}_0^{-1}$.
It is specific for regression with Gaussian prior
that, given ${\bf K}_0^{-1}$, only an
$n$--dimensional equation has to be solved to obtain $h^*$.
)
As the resulting maximum posterior solution 
$p(x|c,h^*)$ is Gaussian by definition,
we find for its mean
\be
\int \!dx \,x\, p(x|c,h^*) 
= h^*(c)
.
\ee
It is not difficult to check that, 
for regression with a Gaussian prior,
$h^*(c)$ is also equal to the mean
$\int \!dx\, x\, p(x|c,D)$
of the exact predictive density
(\ref{predictive}).
Furthermore it can be shown that,
in order to minimize the squared error $[x-h(c)]^2$
for (future) test data,
it is optimal to predict outcome $x$ = $h(c)$
for situation $c$.

In the following sections we will apply the Bayesian formalism
to quantum theory.
Hence, training data $x_i$ will represent the results of 
measurements on quantum systems
and conditions $c_i$ will describe the kind of measurements performed.
Being interested in the determination of quantum potentials 
our hypotheses $h$ will in the following represent potentials $v$.

\section{Inverse quantum statistics}
\label{InvQS}

\subsection{The likelihood model of quantum theory}
\label{likelihood-model}

\subsubsection{Measurements in quantum theory}

The state of a quantum mechanical system
is characterized by its density operator $\rho$.
In particular, 
the probability of measuring value $x$ for observable $O$
in a state described by $\rho$
is known to be
\cite{Messiah-1961,Balian-1991}
\be
p(x|O,\rho) 
= {\rm Tr} \Big(P_O(x) \, \rho  \Big)
.
\ee
This defines
the likelihood model of quantum theory.
The observable $O$, represented by a hermitian operator, 
corresponds to the condition $c$ of the previous section.
The projector $P_O(x)$ = $\sum_l\sep{x_l}{x_l}$
projects on the space of eigenfunctions
$\ket{x_l}$ of $O$
with eigenvalue $x$, i.e.,  for which $O\ket{x_l}$ = $x\ket{x_l}$.
For non--degenerate eigenvalues
$P_O(x)$ = $\sep{x}{x}$.

To be specific, we will consider 
the measurement of particle positions,
i.e., the case
$O$ = $\hat x$ with $\hat x$ being the multiplication operator
in coordinate space.
However, the formalism we will develop does not depend on the 
particular kind of measured observable.
It would be possible, for example, to mix 
measurements of position and momentum
(see, for example, Section \ref{ave-ener}).

For the sake of simplicity, we will assume that no {\it classical noise}
is added by the measurement process.
It is straightforward, however, to include
a classical noise factor in the likelihood function.
If, for example, the classical noise
is, conditioned on $x_i$, 
independent of quantum system
then 
\be
p(\bar x_i|O,\rho)
=
\int \! dx_i \, p(\bar x_i|x_i) \, p(x_i|O,\rho)
,
\ee
where we denoted the `true' coordinates by $x_i$
and the corresponding noisy output 
by $\bar x_i$. 
A simple model for
$p(\bar x_i|x_i)$ could be a Gaussian.

In contrast to the (ideal) measurement of a classical system, 
the measurement of a quantum system
changes the state of the system.
In particular, the measurement process acts as projection $P_O(x)$
to the space of eigenfunctions of operator $O$ 
with eigenvalues consistent with the measurement result.
Thus, performing multiple measurements 
under the assumption of a constant density operator $\rho$
requires special care to ensure the correct preparation
of the quantum system before each measurement. 
In particular, considering a quantum statistical system at finite temperature,
as we will do in the the next section,
the time between two consecutive measurements should be large enough
to allow thermalization of the system.

\subsubsection{Likelihood in the canonical ensemble}

From now on we will
consider a quantum mechanical canonical ensemble at temperature $1/\beta$.
Such a system is described by a density operator 
\be
\rho = \frac{e^{-\beta H}}{{\rm Tr} \, e^{-\beta H} }
,
\label{canonical-rho}
\ee
$H$ denoting the Hamiltonian of the system.
Specifically, we will focus on repeated measurements
of a single particle in a heat bath of temperature $1/\beta$
with sufficient time between measurements
to allow the heat bath to reestablish the canonical density operator.
For (non--degenerated) 
particle coordinates $x_i$
the likelihood for $\rho$ 
becomes the thermal expectation
\be
p(x_i|\hat x,\rho) 
={\rm Tr} \left( P_{\hat x} (x_i) \rho \right)
=\sum_\alpha p_\alpha |\phi_\alpha(x_i)|^2 
=<|\phi (x_i)|^2>
,
\ee
$<\cdots>$ denoting the thermal expectation with probabilities
\be
p_\alpha = \frac{e^{-\beta E_\alpha}}{Z},
\quad Z= \sum_\alpha e^{-\beta E_\alpha}
,
\ee
and energies and orthonormalized eigenstates
\be
H\ket{\phi_\alpha}=E_\alpha\ket{\phi_\alpha}
.
\label{eigenEqH}
\ee
In particular, we will consider a hermitian Hamiltonian of the
standard form
$H$ = $T + V$,
with kinetic energy $T$,
being $1/(2m)$ times the negative Laplacian $-\Laplace$ 
for a particle with mass $m$ 
(setting $\hbar$ = $1$),
and local potential
$V(x,x^\prime)$ = $v(x) \delta (x-x^\prime )$.
Thus, in one dimension
\be
H(x,x^\prime)=
\delta (x-x^\prime )\left(-\frac{1}{2m}\frac{\partial^2}{\partial x^2}
+v(x) \right)
,
\ee
where the $\delta$--functional is usually skipped.

For $n$ independent position measurements $x_i$ 
the likelihood for $\rho(v)$, and thus for $v$, becomes
(writing now $p(x_i|\hat x,\rho)$ = $p(x_i|\hat x,v)$)
\be
p(x_T|\hat x,v) 
=\prod_{i=1}^n p(x_i|\hat x,v)
=\prod_{i=1}^n <|\phi(x_i)|^2> .
\label{max-likeli}
\ee
We remark that it is straightforward to allow
$\beta$ to vary between measurements.

\subsubsection{Maximum likelihood approximation}
\label{mla}

A maximum likelihood approach selects
the potential $v$ with
maximal likelihood $p(x_T|\hat x,v)$
under the training data.
Beginning with a discussion of the {\it parametric} approach
we consider a potential 
$v(\xi,x)$ parameterized by a parameter vector $\xi$
with components $\xi_l$.
To find the parameter vector which maximizes the 
training likelihood
we have to solve the stationarity equation
\be
0= \partial_{\xi} p(x_T|\hat x,v)
,
\label{max-likeli-stat}
\ee
$\partial_{\xi}$
=
$\partial/\partial \xi$ denoting the gradient operator
with components $\partial_{\xi_l}$.
Obtaining from Eq.\ (\ref{max-likeli})
\bea
\partial_{\xi} p(x_i|\hat x,v)
&=&
<\big(\partial_{\xi} \phi^*(x_i)\big)\phi (x_i)>
+<\phi^*(x_i) \big(\partial_{\xi} \phi (x_i)\big)>
\label{deriv-likeli}
\\
&-&\beta\left(
<|\phi (x_i)|^2\partial_{\xi} E> 
-<|\phi (x_i)|^2><\partial_{\xi} E > 
\right)
,
\nonumber
\eea
we see that to solve Eq.\ (\ref{max-likeli-stat})
we have to calculate the derivatives of the eigenvalues
$\partial_\xi E_\alpha$ and of 
the eigenfunctions at the data points
$\partial_\xi \phi_\alpha (x_i)$.
Those are implicitly defined by the eigenvalue equation for $H$ = $H(v)$.
To proceed we
take the derivative of the eigenvalue equation 
(\ref{eigenEqH})
\be
\left(\partial_\xi H\right) \ket{\phi_\alpha}
+
H \ket{\partial_\xi \phi_\alpha}
=
\left(\partial_\xi E_\alpha\right) \ket{\phi_\alpha}
+
E_\alpha \ket{\partial_\xi \phi_\alpha}
.
\ee
Projecting onto $\bra{\phi_\alpha}$
yields, using 
$\partial_\xi H$ = $\partial_\xi v$
and the hermitian conjugate of  Eq.\ (\ref{eigenEqH})
we arrive at
\bea
\partial_\xi E_\alpha 
&=&
\frac{\mel{\phi_\alpha}{\partial_\xi v}{\phi_\alpha}}
     {\scp{\phi_\alpha}{\phi_\alpha}}
,
\label{param1}
\\
(E_\alpha-H)
\ket{\partial_\xi \phi_\alpha}
&=&
\left(  \partial_\xi v -\partial_\xi E_\alpha \right)
\ket{\phi_\alpha }
.
\label{param2}
\eea
Because all orbitals with energy $E_\alpha$
(which may be more than one if $E_\alpha$ is degenerated)
are in the null space of the operator $(E_\alpha-H)$,
Eq.\ (\ref{param2}) alone does not determine
$\partial_\xi \phi_\alpha$ uniquely.
We also notice, that because the 
left hand side of Eq.\ (\ref{param2}) vanishes if projected
on a eigenfunction $\phi_\gamma$
with $E_\gamma$ = $E_\alpha$ 
we find for degenerate eigenfunctions
$\mel{\phi_\gamma}{\partial_\xi H}{\phi_\alpha}$ = 0,
if we choose $\scp{\phi_\alpha}{\phi_\gamma}$ = $\delta_{\alpha,\gamma}$.
A unique solution for $\partial_\xi \phi_\alpha$ can be obtained
be setting $\scp{\phi_\gamma}{\partial_\xi \phi_\alpha}$ = 0
for eigenfunctions $\phi_\gamma$ with $E_\alpha$ = $E_\gamma$.
This corresponds to fixing normalization and phase of eigenfunctions
and, in case of degenerate eigenvalues, uses the freedom to 
work with arbitrary, orthonormal linear combinations
of the corresponding eigenfunctions.
Because the operator $(E_\alpha-H)$ is invertible in the space
spanned by all eigenfunctions $\phi_\gamma$ with 
different energy $E_\gamma\ne E_\alpha$, this yields,
using orthonormal eigenfunctions,
\be
\ket{\partial_\xi \phi_\alpha}
=
\sum_{\gamma\atop E_\gamma\ne E_\alpha} \frac{1}{E_\alpha-E_\gamma}
     \ket{\phi_\gamma}\mel{\phi_\gamma}{\partial_\xi v}{\phi_\alpha}
.
\label{inverted}
\ee
For nondegenerated energies the sum becomes $\sum_{\gamma\ne \alpha}$.
The stationarity equation (\ref{max-likeli-stat})
can now be solved iteratively by
starting from an initial guess $v^0$ for $v$,
calculating $E_\alpha(v)$ and $\phi_\alpha(v)$
to obtain 
$\partial_\xi E_\alpha$ and
$\partial_\xi \phi_\alpha(x_i)$
from Eqs.\ (\ref{param1},\ref{param2})
and thus $\partial_\xi p(x_i|\hat x,v)$
from Eq.\ (\ref{deriv-likeli}).
Then a new guess for $v$ is calculated 
(switching to log--likelihoods)
\be
v^{\rm new} = v^{\rm old}+\eta A^{-1}
     \sum_i \partial_\xi \ln p(x_i|\hat x,v^{\rm old}) 
,                
\ee
with some step width $\eta$ and some positive definite
operator $A$ (approximating for example the Hessian of $\ln p(x_T|\hat x,v)$).
This procedure is now iterated till convergence.

While a parametric approach restricts the space of possible potentials $v$,
a {\it nonparametric} approach
treats each function value $v(x)$ itself 
as individual degree of freedom,
not restricting the space of potentials. 
The corresponding nonparametric stationarity equation is obtained 
analogous to
the parametric stationarity equation (\ref{max-likeli-stat}) 
replacing
partial derivatives $\partial_\xi$ with the functional derivative operator
$\delta_v$ = $\delta/\delta v$
with components
$\delta_{v(x)}$ = $\delta/\left(\delta v(x)\right)$
\cite{Choquet-Bruhat--DeWitt-Morette--Dillard-Bleick-1982}.
Because the functional derivative of $H$ is simply
\be
\delta_{v(x)} H (x^\prime,x^{\prime\prime})
= \delta_{v(x)} V (x^\prime,x^{\prime\prime})
= \delta 
(x-x^\prime) \delta (x^\prime-x^{\prime\prime})
,
\label{ev-eqs}
\ee
we get, using the same arguments leading to Eq.\ (\ref{inverted})
\bea
\delta_{v(x)} E_\alpha 
&=& 
\frac{\mel{\phi_\alpha}{\delta_{v(x)} H}
     {\phi_\alpha}}{\scp{\phi_\alpha}{\phi_\alpha}} = 
|\phi_\alpha(x)|^2
,
\label{deltaE-nonp}
\\
\delta_{v(x)} \phi_\alpha(x^{\prime})
&=&
\sum_{\gamma\atop E_\gamma\ne E_\alpha} \frac{1}{E_\alpha-E_\gamma} 
\,
\phi_\gamma(x^{\prime})
\phi^*_\gamma(x) \phi_\alpha (x)
,
\eea
and therefore
\bea
\delta_{v(x)}p(x_i|\hat x,v)
&=&
<\left(\delta_{v(x)}\phi^*(x_i)\right) \phi (x_i)> 
+<\phi^*(x_i)\delta_{v(x)}\phi (x_i)> 
\\&-&
\beta \left(
< |\phi (x_i)|^2 |\phi (x)|^2 >
-< |\phi (x_i)|^2>< |\phi (x)|^2>
\right)
.
\nonumber
\eea
(The partial derivative with respect to parameters $\xi$ and the
functional derivative with respect to $v(x)$ are related 
according to the chain rule
$\partial_{\xi_l} p(x_i|\hat x,v)$
=
$\int\!dx\, (\partial_{\xi_l} v(x)) 
\, \delta_{v(x)} p(x_i|\hat x,v)$
= 
${\bf v}_\xi \delta_{v} p(x_i|\hat x,v)$
with operator
${\bf v}_\xi (l,x)$
= 
$\partial_{\xi_l} v(\xi,x)$.)

The large flexibility of the nonparametric approach
allows an optimal adaption of $v$ to the available training data.
However, as it is well known in the context of learning
it is the same flexibility
which makes a satisfactory generalization 
to non--training data (e.g., in the future) 
impossible,
leading, for example, to `pathological',
$\delta$-functional like solutions.
Nonparametric approaches 
require therefore additional restrictions in form 
of {\it a priori} information.
In the next section we will include 
{\it a priori} information in form of stochastic processes,
similarly to Bayesian statistics with Gauss\-ian processes
[16, 37, 44, 48, 60--63] 
or to classical regularization theory 
\cite{Tikhonov-1963,Vapnik-1982,Wahba-1990}.
In particular, {\it a priori} information will be implemented
{\it explicitly}, by which we mean 
it will be expressed directly in terms of the function values $v(x)$ itself.
This is a great advantage over parametric methods
where {\it a priori} information
is {\it implicit} in the chosen parameterization,
thus typically difficult or impossible to analyze
and not easily adapted to the situation under study.
Indeed, because it is only {\it a priori} knowledge
which relates training to non--training data,
its explicit control is essential
for any successful learning.

\subsection{Prior models for potentials}
\label{prior-models}
\subsubsection{The need for {\it a priori} information}
\label{need-for-prior} 

Typical results of inverse spectral theory show that, for example,
a one--dimensional local potential can be reconstructed
if a set of two complete spectra 
$\{E^{(1)}_\alpha\}_1^\infty$, $\{E^{(2)}_\alpha\}_1^\infty$
is given for two different boundary conditions for $\phi_\alpha$
\cite{Kirsch-1996,Chadan-Colton-Paivarinta-Rundell-1997}.
Alternatively, a single spectrum is sufficient, if 
either a complete set of norming constants 
$u_\alpha$ = $\int \phi_\alpha dx$ is given 
(for a certain normalization of $\phi_\alpha$
which fixes the values of $\phi_\alpha$ on the boundary)
\cite{Gelfand-Levitan-1951}
or the potential is already known on half of the interval
\cite{Hochstadt-Lieberman-1976}.
Results from inverse scattering theory
show under which circumstances a potential can be reconstructed
from, e.g., a complete spectrum and the phase shifts as function of energy
\cite{Chadan-Colton-Paivarinta-Rundell-1997,Newton-1989,Chadan-Sabatier-1989}.
In practice, however, the number of actual measurements
can only be finite.
Thus, even if noiseless measurement devices would be available,
an empirical determination
of a complete spectrum, or of phase shifts as function of energy,
is impossible.
Therefore, to reconstruct a potential 
from experimental data in practice, 
inverse spectral or inverse scattering theory
has to be combined with additional {\it a priori} information.
If such {\it a priori} information is not made explicit
--- as we try to do in the following --- 
it nevertheless enters any algorithm at least implicitly.

We address in this paper
the measurement of arbitrary quantum mechanical observables,
not restricted to spectral or scattering data.
In particular, we have considered 
the measurement of particle positions.
However,
measuring particle positions only
can usually not determine a quantum mechanical potential completely.
For example, consider the ideal case of an
infinite data limit $n\rightarrow\infty$
for a discrete $x$ variable
(so derivatives with respect to $x$ have to be understood as differences)
at zero temperature (i.e., $\beta\rightarrow\infty$).
This, at least, would allow to obtain
$p(x|\hat x,v)$ = $|\phi_0(x)|^2$
to any desired precision.
But even when we restrict to the case of a local potential, 
we would also need, for example, the ground state energy $E_0$ 
and $\phi_0^*(x)\phi^{\prime\prime}_0(x)$
to determine $v(x)$ from the eigenvalue equation of $H$
\be
v(x) = E_0+\frac{1}{2m}
\frac{\phi_0^*(x)\phi^{\prime\prime}_0(x)}{|\phi_0(x)|^2}
,
\ee
where 
$\phi_0^{\prime\prime}$
=
$\partial^2\phi_\alpha(x)/\partial x^2$
(or a discretized version thereof).
For finite data, a nonlocal potential, continuous $x$,
or finite temperature the situation is obviously even worse.
In the high temperature limit, for example,
$p(x|\hat x,v)$ becomes uniform
and independent from the potential.
Summarizing, even in the ideal case where the complete 
true likelihood $p(x|\hat x,v)$ is assumed to be known, 
the problem of reconstructing potential can still be ill--posed.
(The corresponding time--dependent problem, 
i.e., the reconstruction of a potential $v$ 
given the complete time--dependent likelihood,
is treated in  
\cite{Zhu-Rabitz-1999}. 
A Bayesian approach for time--dependent systems,
based on finite data, 
can be found in \cite{Lemm-TDQ-2000}.)
Hence, while {\it a priori} information is crucial for every learning problem
\cite{Lemm-1996,Lemm-1998,Lemm-1998b},
the reconstruction of a quantum mechanical potential
is particularly sensitive
to the implemented {\it a priori} information.

\subsubsection{Gaussian processes and smooth potentials}
\label{smooth-potentials}

In this section we include, in addition to the likelihood terms,
{\it a priori} information in form of a prior density $p_0(v)$. 
Having specified $p_0(v)$ a Bayesian approach aims at
calculating the predictive density (\ref{predictive}).
Within a maximum posterior approximation 
the functional integral in Eq.\ (\ref{predictive})
can be calculated by Monte Carlo methods
or, as we will do in the following, 
in saddle point approximation, i.e., by selecting
the potential with maximal posterior.
The posterior density of $v$ is 
according to Eq.\ (\ref{bayes-rule}) proportional to the product
of training likelihood and prior
\be
p(v|D) \propto p_0(v) \prod_i <|\phi (x_i)|^2> 
\label{posterior-prob}
.
\ee
Hence, the maximum likelihood approximation 
we have discussed in the last section
is equivalent to a maximum posterior approximation 
under the assumption of a uniform prior.

Technically the most convenient priors are 
{\it Gaussian processes}
which we already have introduced in
(\ref{gaussprior1}) for regression models.
Such priors read for $v$,
\be
p_0(v) 
=
\left(\det \frac{\lambda}{2\pi}{\bf K}_0 \right)^\frac{1}{2}
e^{-\frac{\lambda}{2} 
\mel{v-v_0}{{\bf K}_0}{v-v_0}}
,
\label{gaussprior}
\ee
with mean $v_0$, representing a 
reference potential or template for $v$,
and real symmetric, positive (semi--)definite
covariance operator $(1/\lambda){\bf K}_0^{-1}$,
acting on potentials $v$ and not on wave functions $\phi_\alpha$.
The operator ${\bf K}_0$ defines a scalar product and thus a distance 
measuring the deviation of $v$ from $v_0$.
The most common priors are {\it smoothness priors}
where ${\bf K}_0$ is taken as differential operator.
(In that case ${\bf K}_0$ defines a Sobolev distance.)
Examples of smoothness related inverse prior covariances 
are the negative Laplacian
${\bf K}_0$ = $-{\Laplace}$,
which we have already met in Eq. (\ref{smooth1}),
or operators with higher derivatives
like a ``Radial Basis Functions'' prior
with pseudo-differential operator
${\bf K}_0$ = $\exp{(-{\sigma_{\rm RBF}^2}{\Laplace}/2)}$.

Finally, we want to mention that
also the prior density can be parameterized,
making it more flexible.
Parameters of the prior density, also known as {\it hyperparameters},
are in a Bayesian framework
included as integration variables in Eq.\ (\ref{predictive}), 
or, in maximum posterior approximation,
in the maximization of Eq.\ (\ref{map-def})
\cite{Bishop-1995b,Berger-1980}.
Hyperparameters allow to transform
the point--like maxima of Gaussian priors
to submanifolds of optimal solutions.
For a Gaussian process prior, for example, 
the mean or reference potential $v_0$ and the covariance
${\bf K}_0^{-1}/\lambda$ can so be adapted to the data
\cite{Lemm-1996}.

\subsubsection{Approximate symmetries}

To be more general let us consider
{\it a priori} information related to some {\it approximate symmetry}
\cite{Lemm-1998b}.
In contrast to an exact symmetry 
where it is sufficient to restrict $v$ to be symmetric,
approximate symmetries require
the definition of a distance measuring the deviation from exact symmetry.
In particular, consider 
a unitary symmetry operation $S$, i.e., $S^\dagger=S^{-1}$,
$S^\dagger$ denoting the hermitian conjugate of $S$.
Further, define an operator ${\bf S}$
acting on (local or nonlocal) potentials $V$, by
${\bf S} V$ = $S^\dagger V S$. 
In case of an exact symmetry $V$ commutes with $S$,
i.e., $[V,S]$ = $0$
and thus ${\bf S} V$ = $S^\dagger V S$ = $V$.
In case of an approximate symmetry
we may choose a prior 
\be
p_0(V) \propto e^{-E_S}
,
\ee
with `symmetry energy' or `symmetry error'
\be
E_S
=\frac{1}{2}\mel{V-{\bf S}V}{{\bf K}_S}{V-{\bf S}V}
=\frac{1}{2}\mel{V}{{\bf K}_0}{V}
,
\label{symm-prior}
\ee
some positive (semi--)definite ${\bf K}_S$,
hence positive semi--definite
${\bf K}_0$ 
= 
$({\bf I}-{\bf S})^\dagger {\bf K}_S ({\bf I}-{\bf S})$,
${\bf I}$ denoting the identity.
(Symmetric $V$ are within the Null space of ${\bf K}_S$.)
If ${\bf S}$ belongs to a Lie group
it can be expressed by a Lie group parameter $\theta$
and the generator ${\bf s}$ 
of the corresponding infinitesimal symmetry operation
as ${\bf S}(\theta)$ = $\exp(\theta {\bf s})$.
Hence, we can define an error with respect to 
the infinitesimal operation ${\bf s}$
with, say, ${{\bf K}_S}$ = ${\bf I}$,
\be
E_s
=\lim_{\theta\rightarrow0}
 \frac{1}{2}\scp{\frac{V-{\bf S}(\theta)V}{\theta}}                
                {\frac{V-{\bf S}(\theta)V}{\theta}}
=\frac{1}{2}\mel{V}{{\bf s}^\dagger {\bf s}}{V}
.
\ee
Choosing, for instance, ${\bf s}$ as the derivative operator
(for vanishing or periodic boundary terms)
results in the typical Laplacian smoothness prior
which measures the degree of symmetry of $v$
under infinitesimal translations.

Another possibility to implement approximate symmetries
is given by 
a prior with symmetric reference potential
$V_S$ = ${\bf S}V_S$ 
\be
E_{V_S}=\frac{1}{2}\scp{V-V_S}{V-V_S}
.
\ee
In contrast to Eq.\ (\ref{symm-prior})
which is minimized by any symmetric $V$,
this term is minimized only by $V$ = $V_S$.
Note, that also in Eq.\ (\ref{symm-prior}) an explicit 
non--zero reference potential $V_0$
can be included, meaning that not $V$ but the difference 
$V-V_0$ is expected to be approximately symmetric.

Finally, a certain deviation $\kappa$ from exact symmetry
might even be expected.
This can be implemented by including
`generalized data terms' \cite{Lemm-1996}
\be
E_{S,\kappa}
=\frac{1}{2}(E_S(V)-\kappa)^2
=\frac{1}{2}(\frac{1}{2}\scp{V-{\bf S}V}{V-{\bf S}V}-\kappa)^2
,
\ee
similar to the usual mean squared error terms used in regression.

\subsubsection{Mixtures of Gaussian process priors}

Stochastic process priors have, compared to priors over parameters $\xi$,
the advantage of implementing {\it a priori} knowledge explicitly
in terms of the function values $v(x)$.
Gaussian processes, in particular, always correspond 
to simple quadratic error surfaces, i.e., concave densities.
Being technically very convenient, this is, 
on the other hand, a strong restriction.
Arbitrary prior processes, however, can easily
be built by using mixtures of Gaussian processes
without loosing the advantage of an explicit prior implementation
\cite{Lemm-1996,Lemm-1998,Lemm-1999,Lemm-1998b}.
(We want 
to point out that using a mixture of Gaussian process priors 
does not restrict $v$ to a mixture of Gaussians.)

A mixture of $M$ Gaussian processes 
with component means $v_k$ and 
inverse component covariances $\lambda{\bf K}_k$
reads
\be
p_0(v) 
=\sum_k^M p(v,k)
=\sum_k^M p(k) \,p_0(v|k)
=\sum_k^M \frac{p(k)}{Z_k}e^{-\frac{\lambda}{2}\mel{v-v_k}{{\bf K}_k}{v-v_k}}
\label{gauss-mixture}
\ee
with $Z_k$ 
=
$\left(\det \frac{\lambda}{2\pi} {\bf K}_k\right)^{-\frac{1}{2}}$
and mixture probabilities $p(k)$.
The parameter $\lambda$ plays the role of an inverse mixture temperature.
Analogous to annealing techniques
changing $\lambda$ allows to control the degree of concavity
of the mixture \cite{Lemm-1998,Lemm-1999}.

\subsubsection{Average energy}
\label{ave-ener}

Using a standard Gaussian smoothness prior as in Eq.\ (\ref{gaussprior})
with zero reference potential
$v_0 \equiv 0$ (and, say, zero boundary conditions for $v$)
leads to flat potentials for large smoothness factors $\lambda$.
Especially in such cases it turned out to be useful
to include besides smoothness also {\it a priori} information
which determines the depth of the potential.
One such possibility is to include information about the {\it average energy}
\be
U =
\sum_\alpha p_\alpha E_\alpha 
=\,
<E>
.
\ee
We may remark, that for fixed boundary values of $v$ 
a certain average energy cannot be obtained 
by simply adding a constant to the potential.
The average energy can, however, be set to a value $\kappa$
by including a Lagrange multiplier term
\be
E_U =
\mu (U - \kappa)
,
\ee
Similarly, and technically sometimes easier,
one can include noisy `energy data' of the form
\be
p_{{}_{\scriptsize U}}  
\propto e^{-E_U}
,\quad
E_U =
\frac{\mu}{2} (U - \kappa)^2
.
\label{averageE-penal}
\ee
For $\mu\rightarrow\infty$
this results in $U\rightarrow\kappa$ 
so both approaches coincide.

\subsubsection{Maximum posterior approximation}

Let us consider prior densities 
being a product of a Gaussian 
prior $p_0$ as in  Eq.\ (\ref{gaussprior}),
or more general a mixture of Gaussian processes as in
Eq.\ (\ref{gauss-mixture}),
and a non--Gaussian energy prior 
$p_{{}_{\scriptsize U}}$  
of the form 
of Eq.\ (\ref{averageE-penal}).
In that case, the stationarity equation we have to solve 
to maximize the posterior density of Eq.\ (\ref{posterior-prob})
reads
\be
0 = 
\delta_{v(x)} \ln p_0(v)
+\delta_{v(x)} \ln p_{{}_{\scriptsize U}}(v)  
+\sum_i \delta_{v(x)} \ln p(x_i|\hat x,v)
.
\label{posterior-statEQ}
\ee
While $\delta_{v(x)} p(x_i|\hat x,v)$
has already been calculated in Sect.\ \ref{mla},
we now need also
$\delta_{v(x)} p_0(v)$ and
$\delta_{v(x)} p_{{}_{\scriptsize U}}(v)$  
For a Gaussian $p_0(v)$ the functional derivative is easily found to be
\be
\delta_{v} \ln p_0 
=
\frac{\delta_{v} p_0}{p_0} 
= -\lambda{\bf K}_0(v-v_0)
.
\ee
Similarly, for a mixture of Gaussian processes 
it is not difficult to show that 
\be
\delta_{v} \ln p_0
=-\lambda \sum_k^M p_0(k|v) {\bf K}_k (v-v_k) 
\ee
where $p_0(k|v)$ = ${p_0(v,k)}/{p_0(v)}$.

To get the functional derivative of the non--Gaussian $p_U$ 
we calculate first
\be
\delta_{v(x)} U =
<\delta_{v(x)} E>
-\beta <E\, \delta_{v(x)} E>
+\beta <E> <\delta_{v(x)} E> 
.
\ee
As $\delta_{v(x)} E_\alpha$ has been found in 
Eq.\ (\ref{deltaE-nonp})
this yields
\be
\delta_{v(x)} E_U =
  \mu\left(U-\kappa\right)
  \left(<|\phi (x)|^2>
  -\beta \left( <E \, |\phi (x)|^2>
  -U <|\phi (x)|^2> 
\right)
\right)
.
\ee

Collecting all terms, we can now solve
the stationarity equation (\ref{posterior-statEQ})
by iteration
\be
v^{\rm new}
 =
v^{\rm old}\! +
\eta {\bf A}^{-1}\left(
\lambda {\bf K}_0 (v_0\!-\!v^{\rm old})
+
\sum_i \delta_v\ln p(x_i|\hat x,v^{\rm old})
-\delta_v E_U
\right)
,
\label{iter1}
\ee
where we introduced
${\bf K}_0$ 
=
$
\sum_k p_0(k|v)\,{\bf K}_k$
and
$v_0$ 
= 
${\bf K}_0^{-1} \sum_k p_0(k|v)\, {\bf K}_k v_k$
and a step width $\eta$ 
and positive definite iteration matrix ${\bf A}$
has to be selected.
Choosing  ${\bf A}$ as the identity matrix
means moving in the direction of the gradient
of the posterior.
Taking for ${\bf A}$ the Hessian
one obtains the Newton method.
Quasi--Newton methods, 
like the 
DFP (Davidon--Fletcher--Powell) or
BFGS (Broyden--Fletcher--Goldfarb--Shanno)
variable metric methods,
approximate
the Hessian iteratively 
[68--72] 
(For the case of solving for continuous functions 
see \cite{Airapetyan-Puzynin-1997}.)

A simple and useful choice in our case is 
${\bf A}$ = ${\bf K}_0$
which approximates the Hessian.
For a single Gaussian prior
this choice does
not depend on $v$ and has thus not to be recalculated during iteration.
Eq.\ (\ref{iter1}) then becomes 
\be
v^{\rm new}
 =
(1-\eta) v^{\rm old}\! +
\eta \left(
v_0\!
+(\lambda {\bf K}_0)^{-1}
\Big(
\sum_i \delta_v\ln p(x_i|\hat x,v^{\rm old})
-\delta_v E_U
\Big)
\right)
\label{iter2}
.
\ee

Due to the nonparametric approach for the potential
combined with {\it a priori} information implemented as stochastic
process,
the Bayesian approach formulated in the previous sections
is clearly computationally demanding.
The situation for inverse quantum theory is worse
than, e.g., for  Gaussian process priors
in regression problems (i.e., for a Gaussian likelihood,
local in the regression function)
where it is only necessary to work
with matrices having
a dimension equal to the number of training data
\cite{Wahba-1990,Williams-Rasmussen-1996}.
In our case, where the likelihood is nonlocal in 
the potential and also non--Gaussian prior terms
may occur, the stationarity equation
has to be solved be discretizing the problem.

The following section demonstrates
that at least one--dimensional problems
can be solved numerically 
without further approximation.
Higher dimensional problems, however, e.g., for many--body systems,
require additional approximations.
In such higher dimensional situations, 
the potential may be parameterized
(without skipping necessarily the prior terms)
or the problem has to be divided
in lower dimensional subproblems, e.g., 
by restricting $v$ to certain
(typically, additive or multiplicative)
combinations of lower dimensional functions.
(Similarly, for example to 
additive models \cite{Hastie-Tibshirani-1990}
projection pursuit \cite{Huber-1985}
or neural network like
\cite{Bishop-1995b}
approaches.)

\subsection{Numerical case studies in  inverse quantum statistics}
\label{numerics}
\subsubsection{Approximately periodic potentials}
\label{periodic}

As a first numerical application we discuss the reconstruction of a 
one--dimensional periodic potential.
For example, such a potential may represent a 
one--dimensional solid surface.
To be specific, assume we expect the potential to be periodic,
or even more, to be similar
to a certain periodic reference potential $v_0$.
However, we do not want to restrict the approximation to a parametric form
but want to keep the approximating potential flexible,
so it can adapt to arbitrary deviations from
the periodic reference potential as indicated by the data. 
For example, such deviations may be caused by impurities on 
an otherwise regular surface.
Assuming the deviations from the reference
to be smooth on a scale defined by $\lambda$,
these assumptions can be implemented by a Gaussian smoothness prior
with mean $v_0$, and, say,  Laplacian inverse covariance.
Including the likelihood terms for the empirical data,
and possibly a term $p_U$ adapting the average energy,
we end up with an error functional (negative log--posterior)
to be minimized
\be
-\ln p(v|D) = -\sum_i \ln p(x_i|\hat x,v)
-\frac{\lambda}{2} \mel{v-v_0}{{\Laplace}}{v-v_0}
+E_U
.
\label{periodicE}
\ee

\begin{figure}
\begin{center}
\epsfig{file=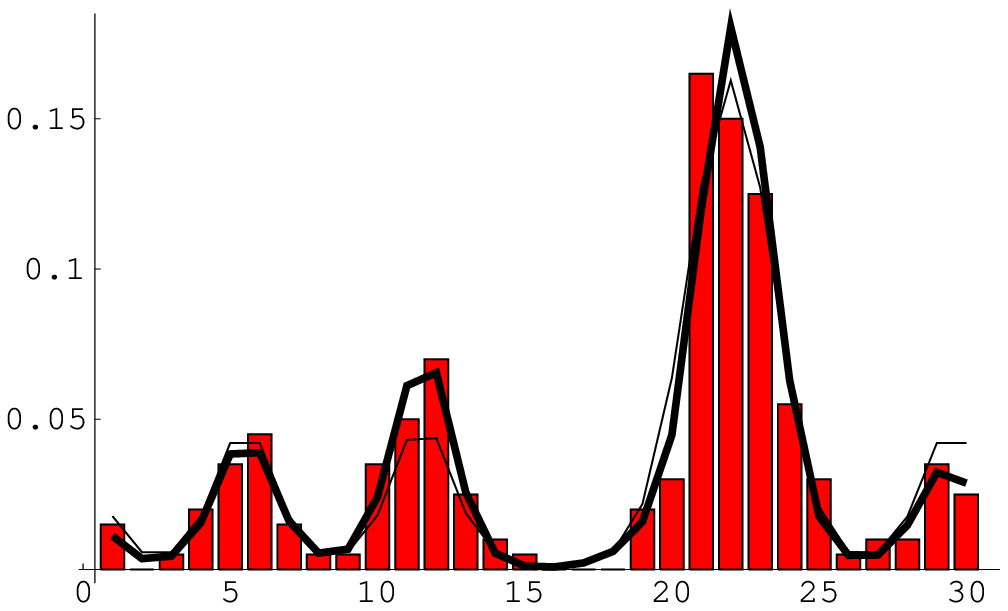, width= 67mm}
$\!\!\!$
\epsfig{file=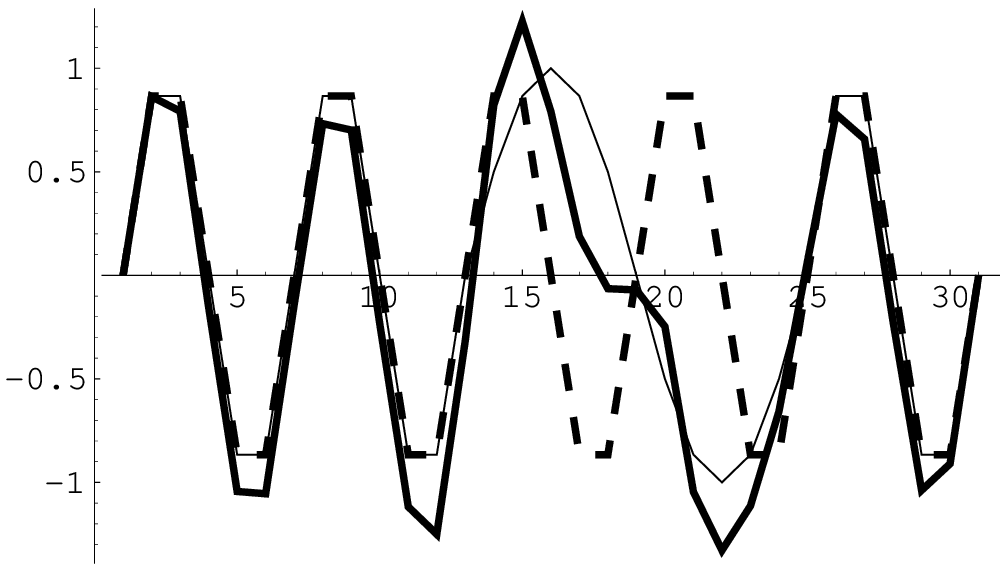, width= 67mm}
\setlength{\unitlength}{1mm}
\begin{picture}(0,0)
\put(-45,40){\makebox(0,0){likelihoods}}
\put(20,40){\makebox(0,0){potentials}}
\put(40.5,40.5){\makebox(0,0){$v_{\rm true}$}}
\put(40,39){\vector(-1,-1){1.5}}  
\put(50.5,30){\makebox(0,0){$v_0$}}
\put(52,7){\makebox(0,0){$v$}}
\end{picture}
\end{center}
\vspace{-0.5cm}
\caption{Reconstruction of an approximately periodic potential
from empirical data.
The left hand side shows likelihoods 
and the right hand side potentials:
Original likelihood and potential $v_{\rm true}$ (thin lines),
approximated likelihood and potential $v$ (thick lines),
empirical density (bars),
reference potential $v_0$ (dashed).
The parameters used are:
200 data points for a particle with mass $m$ = 0.25, 
inverse physical temperature $\beta$ = 4,
inverse Laplacian covariance 
with $\lambda$ = 0.2.
(Average energy $U (v_{\rm true})$ = $-0.3539$
for original potential and
$U(v)$ = $-0.5521$
for the reconstructed potential.)
Notice, that the reconstructed potential $v$
shows clearly the deviation from the strictly periodic
reference potential $v_0$.
}
\label{periodic-fig}
\end{figure}

\begin{figure}
\begin{center}
\epsfig{file=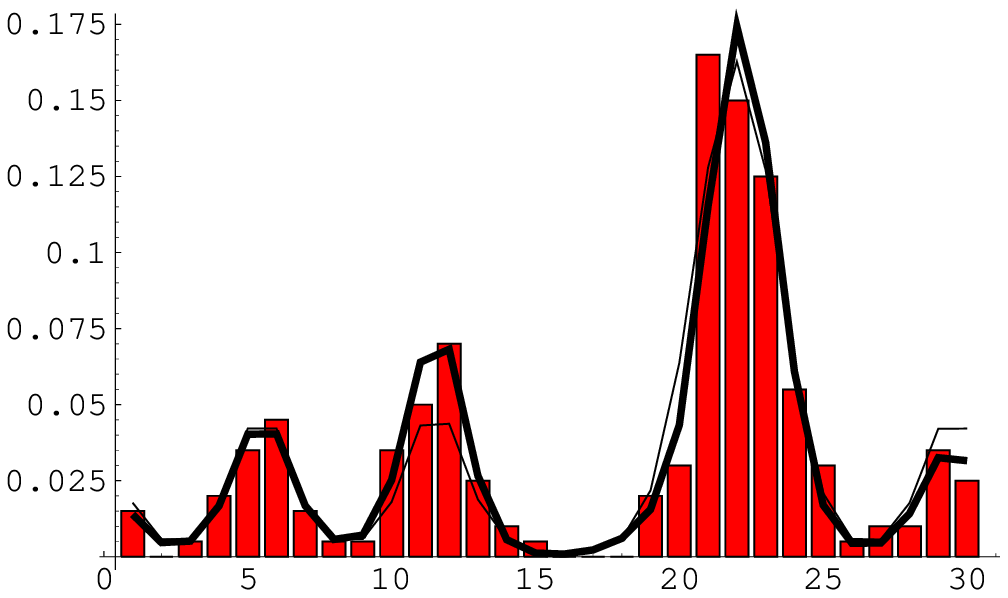, width= 67mm}
$\!\!\!$
\epsfig{file=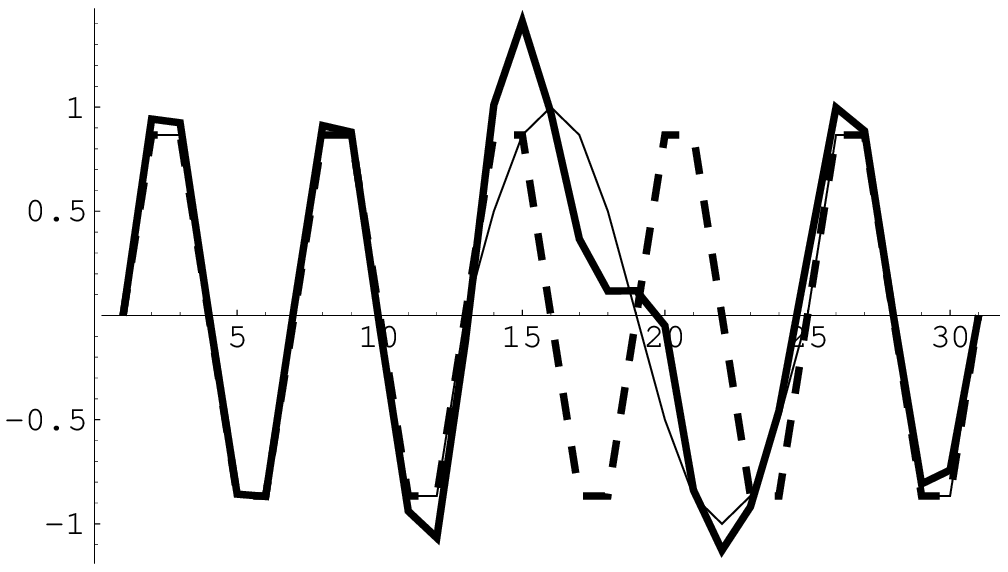, width= 67mm}
\setlength{\unitlength}{1mm}
\begin{picture}(0,0)
\put(-45,40){\makebox(0,0){likelihoods}}
\put(20,40){\makebox(0,0){potentials}}
\put(43,40){\makebox(0,0){$v_{\rm true}$}}
\put(42.5,38){\vector(-1,-2){2.0}}  
\put(50.5,30){\makebox(0,0){$v_0$}}
\put(52,6){\makebox(0,0){$v$}}
\end{picture}
\end{center}
\vspace{-0.5cm}
\caption{
Same data and parameter as for Fig.\ \ref{periodic-fig},
except for a nonzero energy penalty term $E_U$
with $\mu$ = $1000$
and $\kappa$ = $-0.3539$ = $U (v_{\rm true})$.
While there are,
compared to Fig.\ \ref{periodic-fig},
only slight modifications of the likelihood
the average energy
of the reconstructed potential,
$U(v)$ = $-0.3532$, 
is now nearly the same as that 
of the original potential.
}
\label{periodic-fig2}
\end{figure}

Fig.\ \ref{periodic-fig} shows representative examples
of numerical results 
for functional (\ref{periodicE})
without energy penalty term $E_U$ and
with a periodic reference potential (dashed line),
$v_0(x)$ = $\sin(\pi x/3)$,
on a one--dimensional grid with 30 points.
Data have been sampled according to a likelihood
function derived from 
a `true' or original potential $v_{\rm true}$ (shown as thin line)
under periodic boundary conditions for $\phi_\alpha$.
The reconstructed potential (thick line)
has been obtained by minimizing
Eq.\ (\ref{periodicE})
iterating according to Eq.\ (\ref{iter1})
with ${\bf A}$ = $-\lambda{\Laplace}$
and zero boundary conditions for $v$
(so ${\bf A}$ is invertible)
and initial guess $v^{(0)}$ = $v_0$.

Notice, that the distortion of the underlying
`true' potential has been clearly identified.
On the other hand the reconstructed potential coincides well 
with the periodic reference potential
at locations where supported by data.

We want to stress two phenomena
which are typical for the reconstruction of potentials
from empirical data and can also be seen in the figures.
Firstly, the approximation of the likelihood
function is usually better than the approximation
of the potential.
This is due to the fact that quite different potentials 
can produce similar likelihoods.
This emphasizes
the relevance of {\it a priori} information 
for reconstructing potentials.
Secondly, 
especially in low data regions, i.e., at high potentials,
the potential is not well determined.
Thus, empirical data mainly contribute
to the approximation of regions with low potential, 
while {\it a priori} information
becomes especially important in regions 
where the potential is large.
More data will can be obtained for high potential regions 
when the temperature is increased
which spreads the data over a wider area.
At the same time, however, the likelihood becomes more uniform 
at large temperatures, making an identification of $v$ more difficult.

Because the reference potential $v_0$ 
has the same average energy $U$ as the underlying original potential
the results are already reasonable
without energy penalty term $E_U$.
Indeed,
Fig.\ \ref{periodic-fig2} 
shows the relatively small influence of an additional
energy penalty term with quite large $\kappa$
on the likelihood function.
Thus, the approximated probability for empirical data is not much altered.
The presence of an $E_U$ term is better visible for the potential.
In particular its minima
fit now better that of the original.
In the next section, where we will work with a zero reference 
potential $v_0\equiv 0$, the energy penalty term $E_U$
will be more important.

\subsubsection{Inverse two--body problems}

As a second example we study the reconstruction of a two--body potential
by measuring inter--particle distances $x_r$.
Consider the
two--body problem
\be
\frac{P_1^2}{2m_1} + \frac{P_2^2}{2m_2}
+v(x_1-x_2) 
\delta (x_1-x_2-x^\prime_1+x^\prime_2 )
\delta (x_1+x_2-x^\prime_1-x^\prime_2 )
\ee
with single particle momenta $P_i$ = $-i\partial/\partial x_i$.
The problem is transformed to a one--body problem
in the relative coordinates 
in the usual way by introducing 
i.e.,
$x_r$ = $x_1-x_2$,
$P_r$ = $(m_1P_1-m_2P_2)/(m_1+m_2)$,
$x_c$ = $(m_1x_1+m_2x_2)/(m_1+m_2)$,
$P_c$ = $P_1+P_2$, 
$m$ = $(m_1m_2)/(m_1+m_2)$,
and $M$ = $m_1+m_2$ resulting in
\be
\left(\frac{P_r^2}{2m} +v(x_r)\right) \psi_\alpha(x_r) 
= E_\alpha \psi_\alpha(x_r)
.
\label{relative}
\ee
The total energy is additive
$E^{\rm total.}_{\alpha}(P_c)$ 
= 
$E_\alpha+P_c^2/(2M)$
so the thermal probabilities $p^{\rm total}$ factorize
and integrating out the center of mass motion leaves
$p_\alpha$
= ${e^{-\beta E_\alpha}}/{Z}$,
with $E_\alpha$ being the eigenvalues of Eq.\ (\ref{relative}).

\begin{figure}
\begin{center}
\epsfig{file=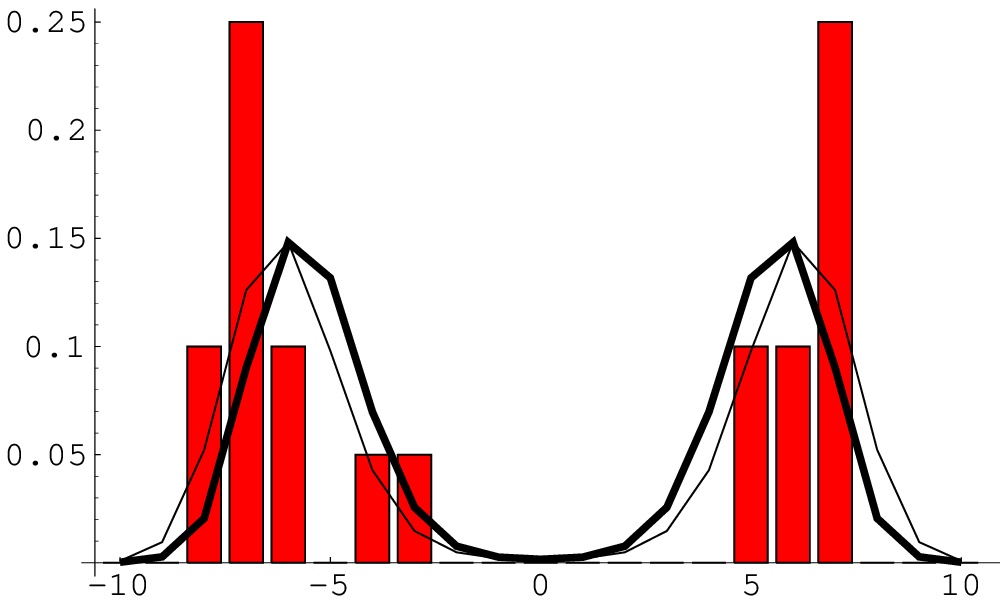, width= 67mm}
$\!\!\!$
\epsfig{file=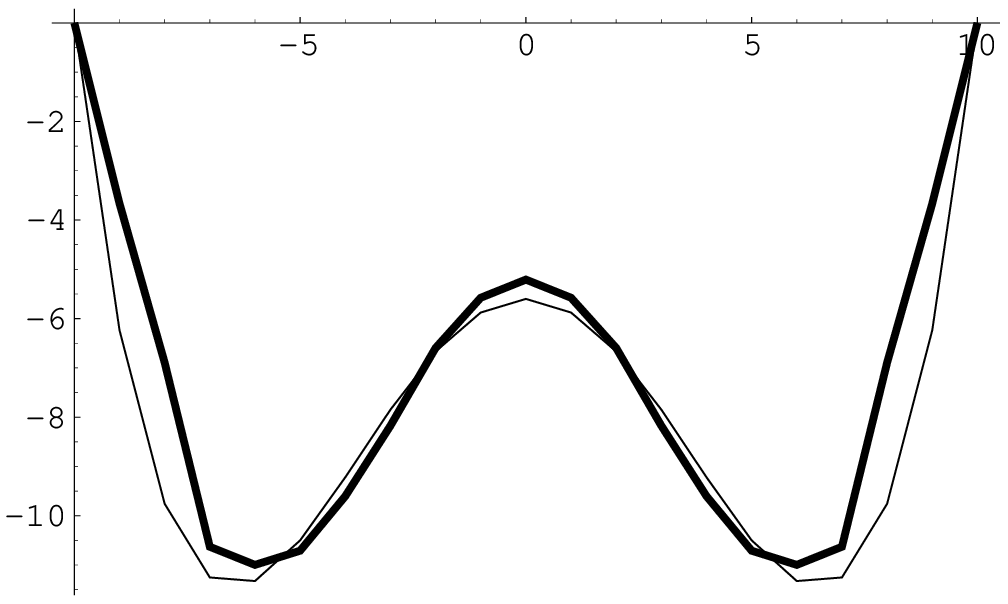, width= 67mm}
\setlength{\unitlength}{1mm}
\begin{picture}(0,0)
\put(-30,39){\makebox(0,0){likelihoods}}
\put(35,39){\makebox(0,0){potentials}}
\put(35,29){\makebox(0,0){$v$}}
\put(35,21){\makebox(0,0){$v_{\rm true}$}}
\end{picture}
\end{center}
\vspace{-0.5cm}
\caption{Approximation of symmetric potential.
Shown are likelihoods (left hand side)
and potentials (right hand side):
Original likelihood and potential (thin lines),
approximated likelihood and potential (thick lines),
empirical density (bars),
The parameters used are:
20 data points for a particle with $m$ = 0.1, 
truncated RBF covariances
as in Eq.\ (\ref{trunc-RBF}) with
$\sigma_{\rm RBF}$ = $7$,
$\lambda$ = $0.001$,
energy penalty term $E_U$ with 
$\mu$ = 20
and reference value $\kappa$ = $-9.66$
= $U(v_{\rm true})$
(average energy $U(v)$ = $-9.33$ for the approximated $v$, 
ground state energy $E_{0}(v)$ = $-9.52$)
inverse physical temperature $\beta$ = 1,
and a potential fulfilling
$v(x)$ = $v(-x)$ and
$v$ = 0 at the boundaries.
}
\label{pot-fig1}
\end{figure}

\begin{figure}
\begin{center}
\epsfig{file=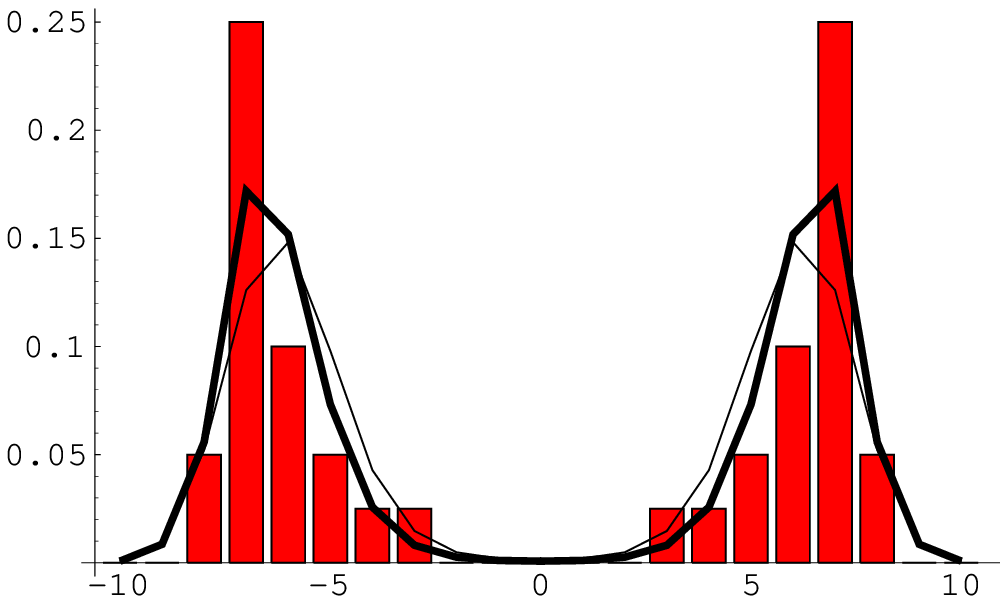, width= 67mm}
$\!\!\!$
\epsfig{file=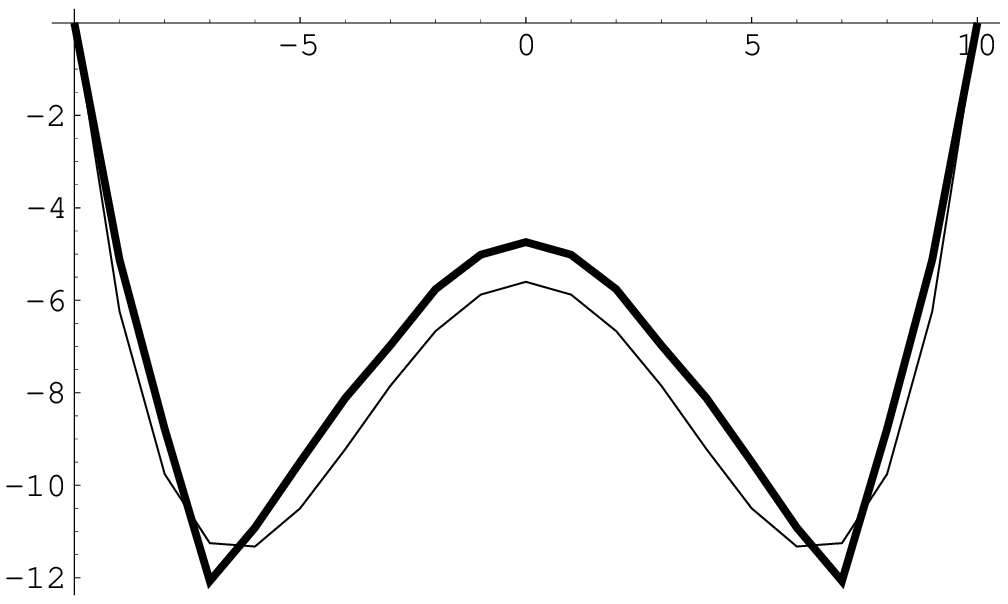, width= 67mm}
\setlength{\unitlength}{1mm}
\begin{picture}(0,0)
\put(-30,39){\makebox(0,0){likelihoods}}
\put(35,39){\makebox(0,0){potentials}}
\put(35,31){\makebox(0,0){$v$}}
\put(35,21){\makebox(0,0){$v_{\rm true}$}}
\end{picture}
\end{center}
\vspace{-0.5cm}
\caption{Same data and parameter 
as for Fig.\ \ref{pot-fig1}
with the exception of  $\sigma_{\rm RBF}$ = $4$,
that means with a smaller smoothness constraint,
and $\mu = 5$.
($U(v)$ = $-9.46$.)
To allow an easier comparison with the reconstructed likelihood
the figure shows the symmetrized empirical density
$P_{\rm sym}$ = $(P_{\rm emp}(x)+P_{\rm emp}(-x))/2$.
}
\label{pot-fig2}
\end{figure}

\begin{figure}
\begin{center}
\epsfig{file=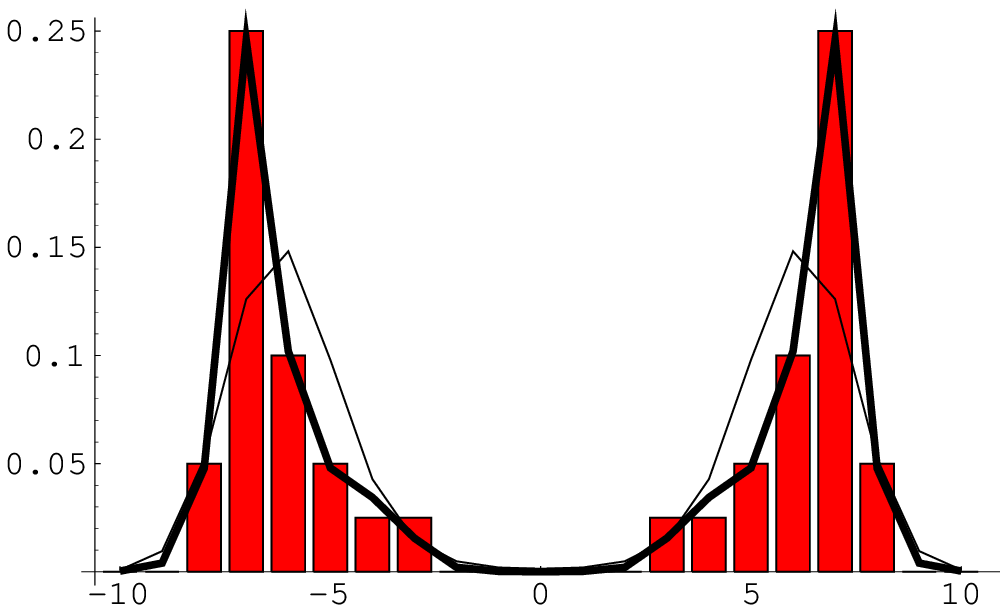, width= 67mm}
$\!\!\!$
\epsfig{file=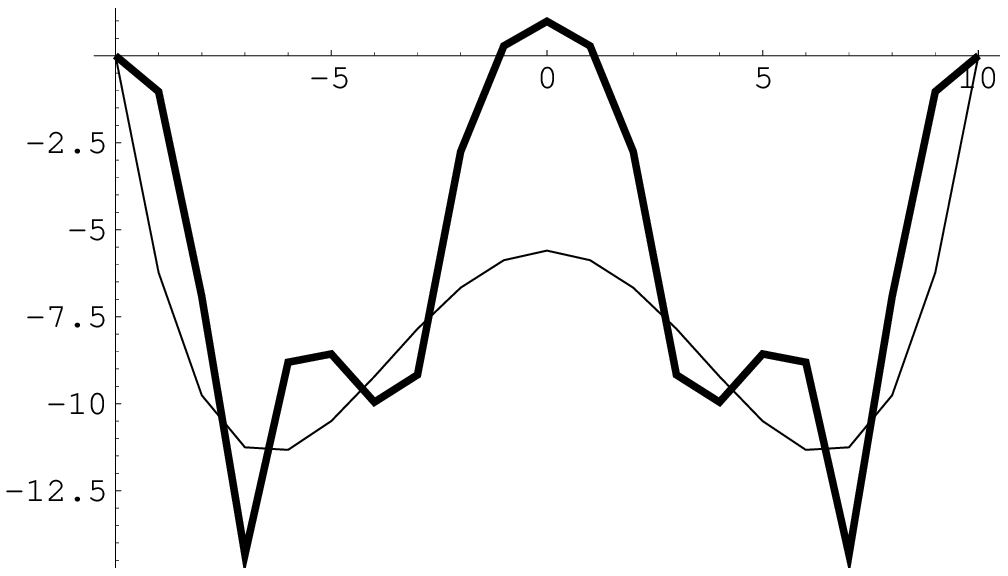, width= 67mm}
\setlength{\unitlength}{1mm}
\begin{picture}(0,0)
\put(-30,39){\makebox(0,0){likelihoods}}
\put(37,10){\makebox(0,0){potentials}}
\put(33,33){\makebox(0,0){$v$}}
\put(36.5,22.5){\makebox(0,0){$v_{\rm true}$}}
\end{picture}
\end{center}
\vspace{-0.5cm}
\caption{Same data and parameter 
as for Fig.\ \ref{pot-fig2}
but with even smaller smoothness constraint
$\sigma_{\rm RBF}$ = $1$.
($\mu = 5$, empirical density symmetrized,
$U(v)$ = $-9.59$.)
Compared with Figs.\ \ref{pot-fig1} and \ref{pot-fig2},
the empirical density is better approximated 
but not the original potential and its likelihood function.
}
\label{pot-fig3}
\end{figure}

Figs.\ \ref{pot-fig1} -- \ref{pot-fig3}
show typical results
for the numerical 
reconstruction of a 
one--dimensional, 
strictly symmetric potential,
fulfilling $v(x)$ = $v(-x)$
and set to zero at the boundaries.
Training data 
have been sampled from a `true' likelihood function 
(thin lines),
resulting in an empirical density 
$P_{\rm emp}(x)$ = $n(x)/n$ (shown by bars), 
where $n(x)$ denotes the number of times the value $x$
occurs in the training data.
The `true' likelihood has been
constructed from a `true' potential (thin lines)
choosing periodic boundary conditions for the wavefunctions.
In contrast to Sect.\ \ref{periodic}
a zero reference potential $v_0\equiv 0$ and
a truncated Radial Basis Function (RBF) prior \cite{Lemm-1996}
has been used 
\be
{\bf K}_0
=
\sum_{k=0}^3 
\frac{\sigma_{\rm RBF}^{2k}}{k!2^k} (-1)^k {\Laplace}^k
,
\label{trunc-RBF}
\ee
(${\Laplace}^k$ denoting the $k$th iterated Laplacian)
which includes, compared to a Laplacian prior,
higher derivatives, hence producing 
a rounder reconstructed potential
(cmp.\ Sect.\ \ref{smooth-potentials}).
The approximated potentials (thick lines) have been obtained
by iterating Eq.\ (\ref{iter1}), including a term $E_U$
adapting the thermal energy average to that of the original potential.
As iteration matrix we used ${\bf A}$ = $\lambda{\bf K}_0$
together with an adaptive step size $\eta$.
An initial guess for the potential
has been obtained by adding negative 
$\delta$--peaks on the data points (except for data on the boundary), i.e.,
$v^{(0)}$ = $-\sum_i\delta_{x,x_i}$.
The number of iterations necessary to obtain convergence
has been typically between 50 and 100.

Comparing Figs.\ \ref{pot-fig1} -- \ref{pot-fig3}
one sees that a smaller smoothness leads to a better fit
of the empirical density. 
A larger smoothness, on the other hand,
leads to better fit in regions where
smoothness is an adequate prior.
Near the boundaries, however,
where the original is relatively steep,
a higher smoothness leads
to a poorer approximation.
A remedy would be, for example,
an adapted reference potential $v_0$.

\begin{figure}
\begin{center}
\psfig{file=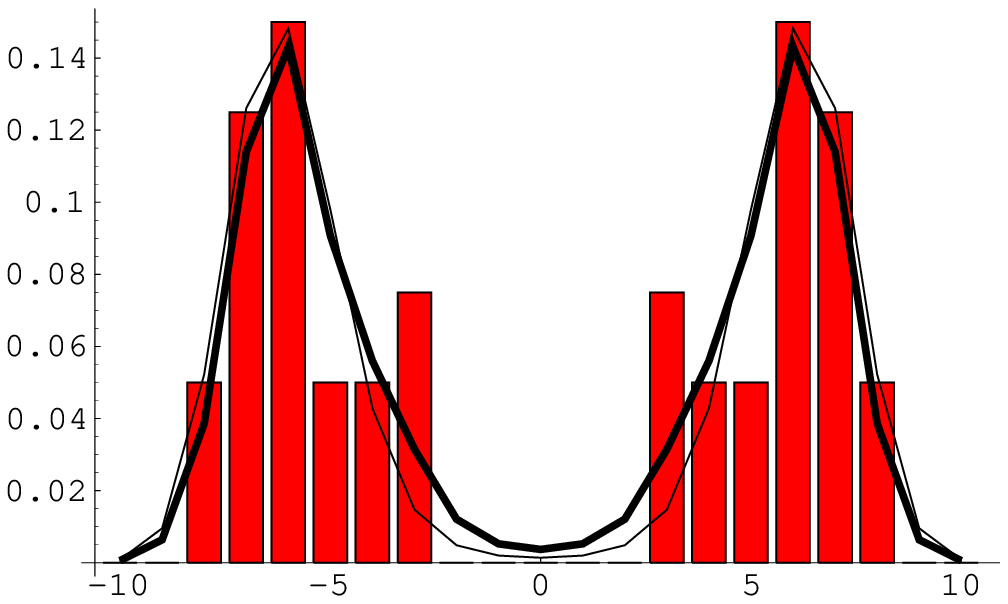, width= 67mm}
$\!\!\!$
\epsfig{file=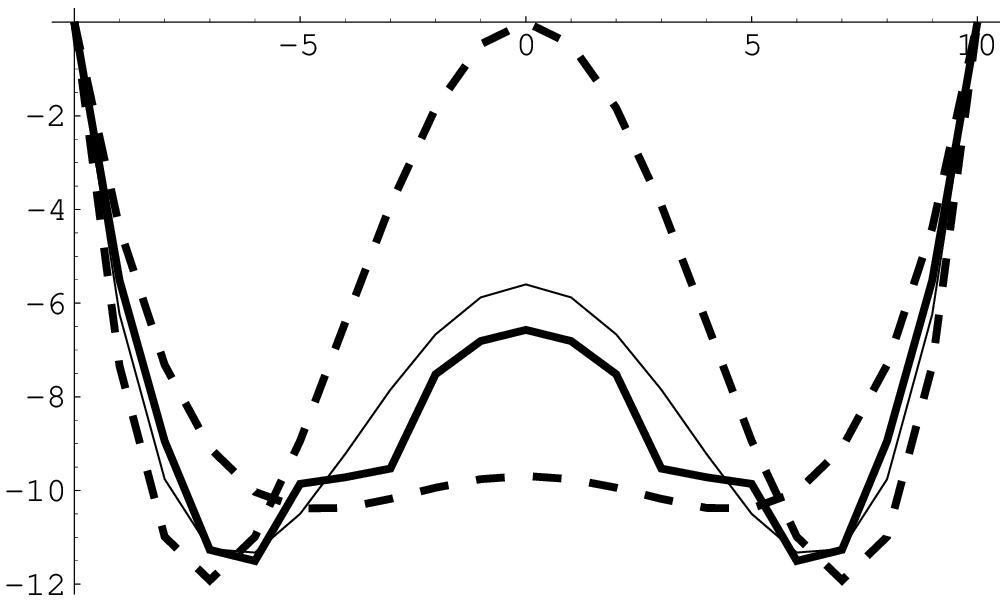, width= 67mm}
\setlength{\unitlength}{1mm}
\begin{picture}(0,0)
\put(-30,39){\makebox(0,0){likelihoods}}
\put(35.5,8){\makebox(0,0){potentials}}
\put(30,33){\makebox(0,0){$v_1$}}
\put(35.5,29){\makebox(0,0){$v_{\rm true}$}}
\put(35.5,21.5){\makebox(0,0){$v$}}
\put(35.5,16){\makebox(0,0){$v_2$}}
\end{picture}
\end{center}
\vspace{-0.5cm}
\caption{Approximation of symmetric potential
with mixture of Gaussian process priors.
The left hand side shows likelihoods 
and the right hand side potentials:
Original likelihood and potential (thin lines),
approximated likelihood and potential (thick lines),
symmetrized
empirical density (bars),
and the two reference potentials $v_1$,$v_2$ 
(dashed, $v_2$ deeper in the middle).
The parameters used are:
20 data points for a particle with $m$ = 0.1,
inverse physical temperature $\beta$ = 1, 
${\bf K}_0$ = $-\Laplace$,
inverse mixture temperature $\lambda$ = 0.1,
energy penalty factor $\mu$ = 10
for average energy $\kappa$ = $-9.66$ = $U(v_{\rm true})$
(and $U(v)$ = $-9.55$, $E_0(v)$ = $-9.82$)
$v(x)$ = $v(-x)$ symmetric, 
and
$v$ = 0 at the boundaries.
Because the data support both reference potentials 
$v_1$ and $v_2$,
the approximated $v$ is in regions with no data essentially
a smoothed mixture between $v_1$ and $v_2$
with mixture coefficients for prior components 
$p_0(1|v)$ = 0.3, $p_0(2|v)$ = 0.7.
}
\label{pot-fig4}
\end{figure}

Fig.\ \ref{pot-fig4}
presents an application of a mixture of Gaussian process priors
as given in Eq.\ (\ref{gauss-mixture}).
Such mixture priors
can in principle be used to construct
an arbitrary prior density,
adapted to the situation under study. 
For the numerical example a two component mixture has been chosen
with equal component variances
${\bf K}_k$ = ${\bf K}_0$ 
of the form of Eq.\ (\ref{trunc-RBF})
and two reference potentials $v_i$ (shown as dashed lines)
with the same average energy $U$.
In the special situation shown in the figure
both reference functions $v_i$ fit similarly well to the empirical data.
(The final mixture coefficients for $v_1$ and $v_2$ 
  are $p_0(1|v)$ = 0.3 and $p_0(2|v)$ = 0.7.)
Hence, in the no--data region
the approximated potential $v$
becomes a smoothed, weighted average of  $v_1$ and $v_2$. 
Because both reference potentials coincide 
also relatively well with the original $v$ near the boundaries,
the approximation in Fig.\ \ref{pot-fig4}
is better than in Figs.\ \ref{pot-fig1} -- \ref{pot-fig3}.

In conclusion, the two one--dimensional examples show 
that a direct numerical solution 
of the presented Bayesian approach to inverse quantum theory
can be feasible.


\subsection{Classical approximation}

Before discussing a possible approximation for many--body systems
we will first study the classical
limit of inverse quantum statistics.
The classical limit is much easier to solve than the full 
quantum mechanical problem
and may, for example for large masses,
already give a useful approximation.
 
The phase space density of a classical canonical ensemble
is given by 
\be
p(x,p_{\rm cl}|v)= Z^{-1}
e^{-\beta\left(\frac{p_{\rm cl}^2}{2m}+v(x)\right)}
,
\ee
with
\be
Z
= \int\! dp_{\rm cl}\,dx\,
e^{-\beta\left(\frac{p_{\rm cl}^2}{2m}+v(x)\right)}
.
\ee
Here we used $p_{\rm cl}$
to denote the classical momentum
to distinguish it from a density $p$.
The probability
$p(x|v)$ for measuring $x$
[to simplify the notation
we abstain in this context from denoting the observable $O$ explicitly]
is then obtained by integrating over $p_{\rm cl}$,
\be
p(x|v)
= \int\! dp_{\rm cl}\,
p(x,p_{\rm cl}|v)
= Z_x^{-1}e^{-\beta v(x)}
,
\label{clLike}
\ee
where
\be
Z_x= \int\! dx\,e^{-\beta v(x)}
.
\ee
Notice, that the classical $p(x|v)$ is mass independent,
and, most important,
that it can be obtained directly from $v(x)$
without having to solve an eigenvalue problem
like in the quantum case.

\begin{figure}
\begin{center}
\epsfig{file=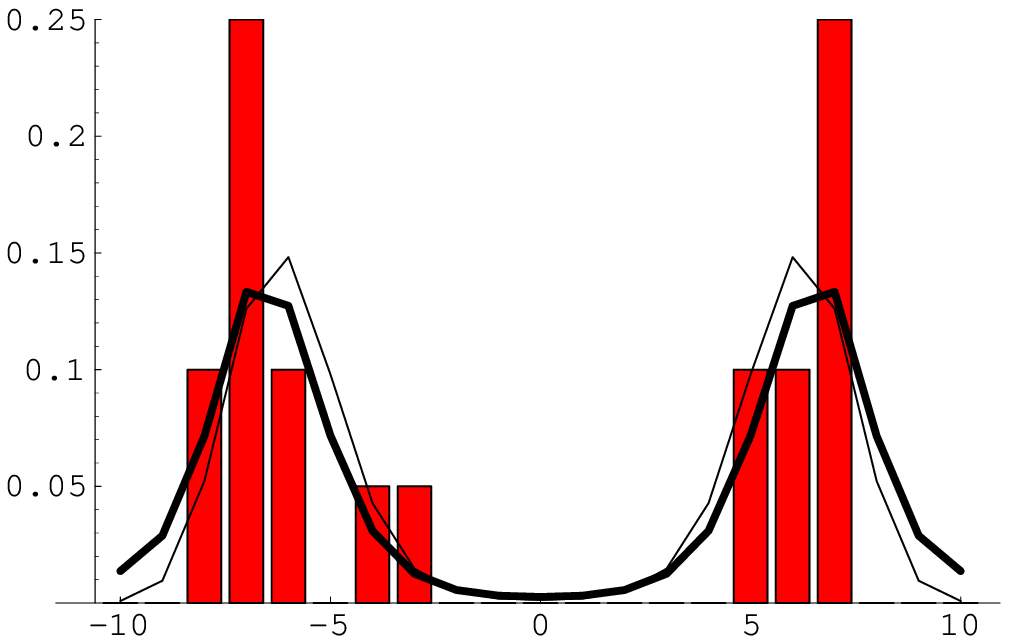, width=67mm}
$\!\!\!$
\epsfig{file=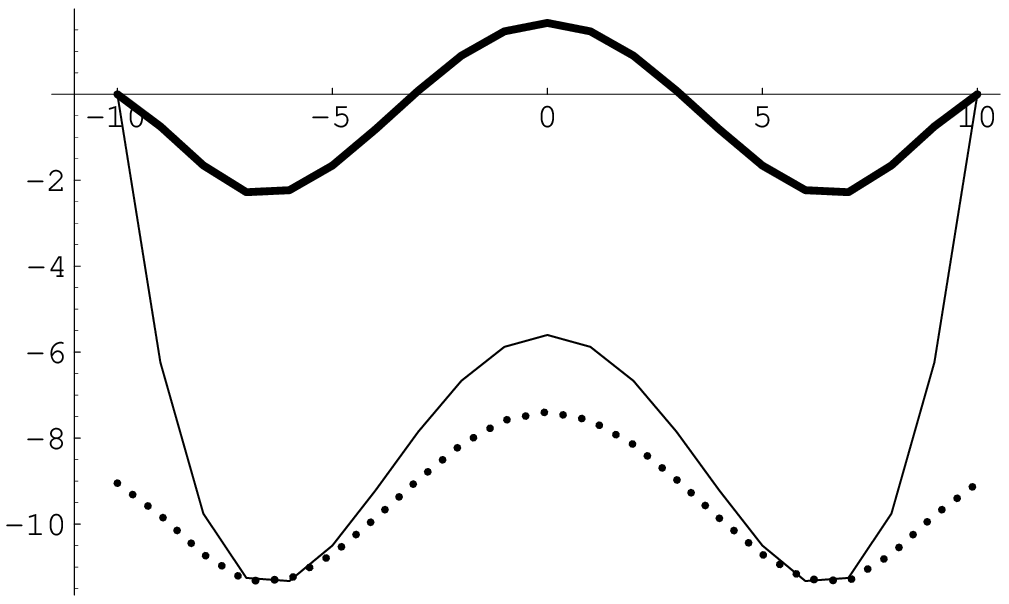, width=67mm}
\setlength{\unitlength}{1mm}
\begin{picture}(0,0)
\put(-30,39){\makebox(0,0){likelihoods}}
\put(36.5,29){\makebox(0,0){potentials}}
\put(27,33.5){\makebox(0,0){$v_{\rm cl}$}}
\put(45,23){\makebox(0,0){$v_{\rm true}$}}
\put(37,12){\makebox(0,0){$v_{\rm cl}-c$}}
\end{picture}
\end{center}
\caption{Classical approximation of symmetric potential.
Shown are likelihoods (left hand side)
and potentials (right hand side):
Original likelihood and potential (thin lines),
approximated likelihood and potential $v_{\rm cl}$ (thick lines),
empirical density (bars).
The dotted line shows 
$v_{\rm cl}-c$ with constant 
$c$ = ${\rm min}[v_{\rm cl}]-{\rm min}[v_{\rm true}]$.
Except for the fact that no
energy penalty term has been used for this classical calculation
the parameters and data are
the same as in Fig.~\ref{pot-fig1}.
(20 data points, sampled from the true quantum mechanical likelihood,
truncated RBF covariances (\ref{trunc-RBF}) with
$\sigma_{\rm RBF}$ = $7$,
$\lambda$ = $0.001$,
inverse physical temperature $\beta$ = 1,
$v(x)$ = $v(-x)$ and
$v$ = 0 at the boundaries.)}
\label{cl-fig1}
\end{figure}

Analogously to the quantum mechanical approach
the classical likelihood model (\ref{clLike})
for position measurements
can now be combined with a prior model for potentials $v$,
leading to a posterior density $p(v|D)$.
In particular, adding a Gaussian process prior 
the log-posterior becomes
\be
\ln p(v|D)
=
-\beta \sum_{i=1}^n v(x_i)
-\frac{\lambda}{2}\mel{v-v_0}{{\bf K}_0}{v-v_0} - n \ln Z_x
-\ln\det\left( \frac{\lambda}{2\pi}{\bf K}_0\right)^{\frac{1}{2}}
.
\ee
Again, we can refer to a maximum posterior approximation
and consider the potential which maximizes the posterior
as the solution of our reconstruction problem.
The corresponding stationarity equation 
is found by setting
the functional derivative of the log-posterior
with respect to $v(x)$ to zero,
\be
0 =
\delta_{v}\ln p(v|D)
=
-\beta N-\lambda{\bf K}_0 (v-v_0) + n \beta p(x|v)
,
\label{clstat}
\ee
where $N$ = $\sum_i \delta (x - x_i)$.
Choosing an initial guess $v^{(0)}$ 
Eq.~(\ref{clstat}) can be solved
by straightforward iteration.
The results of a classical calculation 
(with parameters and data as 
in Fig.~\ref{pot-fig1}, but without energy penalty term)
are shown in Fig.~\ref{cl-fig1}.

\section{Inverse many--body theory}
\label{Inv-many-body}
\subsection{Systems of Fermions}

In this section the Bayesian approach 
for inverse problems will be applied to many--body systems.
To be specific, we will study the simultaneous measurement
of the positions of $N$ particles.
We assume the measurement result to be given
as a vector $x_i$ consisting of $N$ single particle coordinates $x_{i,j}$. 
The treatment can easily be generalized to partial measurements
of $x_i$ by including an integration over components 
which have not been observed.
The likelihood for $v$, if measuring a vector $x_i$ of coordinates, 
becomes for a many--body system
\be
p(x_i|\hat x,v) 
= {\rm Tr} 
\Big(\sep{x_{i,1},\cdots x_{i,N}}{x_{i,1},\cdots x_{i,N}}\, \rho  \Big)
,
\ee
which is now 
a thermal expectation with respect to many--body energies 
$E_\alpha$
\be
p(x_i|\hat x,v) 
= \sum_\alpha p_\alpha |\psi^{(N)}_\alpha(x_{i,1},\cdots ,x_{i,N})|^2 
= <|\psi^{(N)}(x_{i,1},\cdots ,x_{i,N})|^2 >
.
\ee
In particular, we will be interested in {\it fermions}
for which the wave functions $\psi_\alpha$ 
and $\ket{x_{i,1},\cdots ,x_{i,N}}$
have to be antisymmetric.
Considering  a canonical ensemble,
the density operator $\rho$ has still the form 
of Eq.\ (\ref{canonical-rho}),
but with $H$ replaced now by a many--body Hamiltonian.
For fermions, it is convenient to express
the many--body Hamiltonian in second quantization,
i.e., in terms of creation and annihilation operators
\cite{Blaizot-Ripka-1986,Negele-Orland-1988}.
A Hamiltonian with one--body part $T$,
e.g., $T$ = $-(1/2m)\Delta$,
and two--body potential $V$ can so be written
\be
H = T + V 
= \sum_{ij}T_{ij} \,a^\dagger_i a_j
+ \frac{1}{4}\sum_{ijkl} V_{ijkl} \, a^\dagger_i a^\dagger_j a_l a_k 
,
\label{two-boddy-H}
\ee
with antisymmetrized matrix elements $V_{ijkl}$.
Hereby,
$a_\alpha a_\gamma^\dagger +a_\gamma^\dagger a_\alpha $ 
= 
$\scp{\varphi_\alpha}{\varphi_\gamma}$ is equal to the overlap
of the one--body orbitals 
$\ket{\varphi_\alpha}$ = $a_\alpha^\dagger\ket{0}$
and 
$\ket{\varphi_\gamma}$ = $a_\gamma^\dagger\ket{0}$
which are created or destroyed
by the operators $a_\gamma^\dagger$ or $a_\alpha$, respectively.
Furthermore,
$a_\alpha^\dagger a_\gamma^\dagger +a_\gamma^\dagger a_\alpha^\dagger $ 
=0,
$a_\alpha a_\gamma +a_\gamma a_\alpha $ 
=0.
A two--body eigenfunction of the Hamiltonian (\ref{two-boddy-H})
can for example be expanded as follows
\be
\ket{\psi^{(2)}_\alpha}
= 
\sum_{\alpha,\gamma} 
   c_{\alpha,\gamma}
 \ket{\varphi_\alpha,\varphi_\gamma}
,
\ee
where
$\ket{\varphi_\alpha,\varphi_\gamma}$ 
= 
$a_\alpha^\dagger a_\gamma^\dagger\ket{0}$
denotes a Slater determinant
being an antisymmetrized wavefunction.

The symmetrized version of a potential, 
local in relative coordinates, is
\be
V_{x_1 x_2 x^\prime_1 x^\prime_2} 
=v(|x_1-x_2|) \Big(
\delta(x_1-x_1^\prime) \delta(x_2-x_2^\prime)
-
\delta(x_1-x_2^\prime) \delta(x_2-x_1^\prime)
\Big)
.
\label{asymlocpot}
\ee
Here we can always choose $v(0)$ = 0.
Now, assume we are interested in the reconstruction 
of $v(x)$ for $x>0$.
Solving the stationarity equation of the 
maximum posterior approximation
analogous to Eq.\ (\ref{posterior-statEQ}) of Sect.\ \ref{InvQS},
the prior terms remains unchanged
and only the likelihood terms have to be adapted.
Using
\be
\delta_{v(x)} v(|x_1-x_2|) = \delta (x-|x_1-x_2|) 
,
\ee
we find 
\be
\delta_{v(x)} H
=
\frac{1}{2}\sum_{x_1}  
a^\dagger_{x_1} a^\dagger_{x_1-x} 
a_{x_1-x} a_{x_1} 
+
\frac{1}{2}\sum_{x_1}  
a^\dagger_{x_1} a^\dagger_{x_1+x} 
a_{x_1+x} a_{x_1} 
,
\ee
where $x> 0$,
and can write, similar to the one--body case,
\bea
\delta_{v(x)} E_\alpha 
&=&
 \frac{\mel{\psi_\alpha}{\delta_{v(x)} H}{\psi_\alpha}}
             {\scp{\psi_\alpha}{\psi_\alpha}}
,
\label{manybodyVar1}
\\
\ket{\delta_{v(x)} \psi_\alpha}
&=&\sum_{\gamma \atop E_\gamma\ne E_\alpha}
\frac{1}{E_\alpha-E_\gamma} 
\sep{\psi_\gamma}{\psi_\gamma}
\delta_{v(x)} H\ket{\psi_\alpha}
.
\label{manybodyVar2}
\eea
From this the functional derivatives of the likelihoods, 
$\delta_{v(x)}p(x_i|\hat x,v)$,
can be obtained. However, a direct numerical or analytical solution
of the full inverse many--body equations
is usually not feasible.
To deal with this problem, 
a mean field approach 
will be developed 
in the next section.

\subsection{Inverse Hartree--Fock theory}

To tackle the inverse many--body problem
we will treat it in Hartree--Fock approximation
[74--77]. 
Thus, we replace the full many--body Hamiltonian $H$ 
by a one--body Hartree--Fock Hamiltonian 
$H^{HF}$
= $\sum_{kl} h_{kl} a^\dagger_ka_l$
with matrix elements $h$ defined, 
for example in coordinate representation, as
\be
h_{xx^\prime} 
= T_{xx^\prime}  + \sum_k^N \mel{x \varphi_k}{V}{x^\prime \varphi_k}
\label{h-hf}
,
\ee
the $\varphi_k$ being the $N$--lowest 
(orthonormalized) eigenstates of $h$.
The corresponding
eigenvalue equation
\be
h\varphi_k = \epsilon_k \varphi_k
\label{HF-eq}
,
\ee
is nonlinear,
due to the $\varphi_k$--dependent definition (\ref{h-hf}) of $h$,
and has to be solved by iteration.
The Hartree--Fock ground state 
is given by the Slater determinant
$\ket{\Phi_0}$ = $\det\{\ket{\varphi_k}\}$
made from the $N$--lowest orbitals,
and has energy 
$E_0^{HF}$ 
=
$\sum_k^N t_{kk} +\frac{1}{2}\sum_{kl}^N v_{klkl}$
=
$\sum_k^N \epsilon_{k} -\frac{1}{2}\sum_{kl}^N v_{klkl}$.
Considering now the case of zero temperature, 
the many--body likelihood
for the true ground state $\psi_0$,
\be
p(x_i|\hat x,\rho(v))
=
\scp{\psi_0}{x_i}\scp{x_i}{\psi_0}
,
\ee
becomes in Hartree--Fock approximation
\be
p(x_i|\hat x,\rho_{HF}(v))=
\scp{\Phi_0}{x_i}\scp{x_i}{\Phi_0}
.
\ee
The scalar product
of the Hartree--Fock ground state
$\Phi_0$ and the many--body position eigenfunction 
$\ket{x_i}$ corresponding 
to the measured vector $x_i$
is a determinant
and can be expanded in its cofactors $M_{kl;i}$
\be
\scp{x_i}{\Phi_0}
= \det\{\scp{x_{i,l}}{\varphi_k}\}
= \det B_i
= \sum_l^N M_{kl;i} B_{kl;i} 
,
\ee
$B_i$ being the matrix of overlaps
with elements
$B_{kl;i} = \scp{x_{i,l}}{\varphi_k}$
= $\varphi_k(x_{i,l})$.
(For the generalization to non--hermitian $h$ see for example 
\cite{Blaizot-Ripka-1986,Lemm-1995a}.)

To maximize the posterior,
we have to calculate
the functional derivative of the Hartree--Fock likelihood
with respect to the potential
\cite{Lemm-IHF-1999}
\be
\delta_{v(x)} p(x_i|\hat x,\rho_{\rm HF}(v))
= 
\scp{\delta_{v(x)} \Phi_0}{x_i}\scp{x_i}{\Phi_0}
+\scp{\Phi_0}{x_i}\scp{x_i}{\delta_{v(x)} \Phi_0}
.
\ee
Here the factors
\be
\scp{x_i}{\delta_{v(x)} \Phi_0}
= \sum_{kl}^N M_{kl;i} \,\scp{x_{i,l}}{\delta_{v(x)} {\varphi_k}}
= \sum_{kl}^N M_{kl;i} \, \Delta_{kl;i}(x)
,
\ee
can be expressed by single particle derivatives
$\Delta_{kl;i}(x)$ 
= $\scp{x_{i,l}}{\delta_{v(x)} {\varphi_k}}$
= $\delta_{v(x)} {\varphi_k}(x_{i,l})$.
Analogously to Sect.\ \ref{InvQS}
the functional derivatives $\delta_{v(x)} {\varphi_k}$ 
can be obtained from the functional derivative of
Eq.\ (\ref{HF-eq})
\be
(\delta_{v(x)} h) \,\varphi_k + h\, \delta_{v(x)} \varphi_k 
= (\delta_{v(x)} \epsilon_k) \,\varphi_k + \epsilon_k \,\delta_{v(x)} \varphi_k
.
\ee
Projecting onto $\bra{\varphi_k}$
and using the hermitian conjugate of Eq.\ (\ref{HF-eq})
we find the Hartree--Fock version of 
Eqs.\ (\ref{manybodyVar1}) and (\ref{manybodyVar2})
\bea
\delta_{v(x)} \epsilon_k
&=&
\frac{\mel{\varphi_k}{\delta_{v(x)} h}{\varphi_k}}
     {\scp{\varphi_k}{\varphi_k}}
,
\label{delta-eps}
\\
\ket{\delta_{v(x)} \varphi_k}
&=&
\sum_{l\atop \epsilon_l\ne \epsilon_k} 
\frac{1}{\epsilon_k-\epsilon_l} \,
\sep{\varphi_l}{\varphi_l}
\delta_{v(x)} h 
\ket{\varphi_k}
\label{delta-varphi}
,
\eea
where we, as done before, have fixed orthonormalization and phases by
choosing $\scp{\delta_{v(x)}\varphi_k}{\varphi_l}$ = 0 
for orbitals with equal energy.
In contrast to Sect.\ \ref{InvQS}, however,
$h$, and thus $\delta_{v(x)} h$, now obey a nonlinear equation.
Indeed, from Eq.\ (\ref{h-hf}) it follows
\bea
\delta_{v(x)} h_{x^\prime x^{\prime\prime}}
&=& 
\sum_j^N \Big(
\mel{x^{\prime} \,\varphi_j}{\delta_{v(x)}
  V}{x^{\prime\prime}\,\varphi_j}
\label{delta-h}
\\&&
+
\mel{x^{\prime} \,\delta_{v(x)} \varphi_j}{V}{x^{\prime\prime}\,\varphi_j}
+\mel{x^{\prime}\,\varphi_j}{V}{x^{\prime\prime}\,\delta_{v(x)} \varphi_j}
\Big)
\nonumber
.
\eea
Inserting 
Eq.\ (\ref{delta-eps})
and Eq.\ (\ref{delta-h})
into Eq.\ (\ref{delta-varphi}),
we obtain the inverse Hartree--Fock equation 
for $\delta_{v(x)} \varphi_k$ 
\bea
\delta_{v(x)} \varphi_k (x^\prime )
&=&
\sum_{l\atop \epsilon_l\ne \epsilon_k} \frac{1}{\epsilon_k-\epsilon_l}\,
\varphi_l(x^\prime)
\sum_j^N 
\Big(
\mel{\varphi_l\varphi_j}{\delta_{v(x)} V}{\varphi_k\varphi_j}
\label{invHFeq}
\\
&+&
\mel{\varphi_l\delta_{v(x)}\varphi_j}{V}{\varphi_k\varphi_j}
+\mel{\varphi_l\varphi_j}{V}{\varphi_k\delta_{v(x)}\varphi_j}
\Big)
\nonumber
.
\eea
Recalling the definition of the antisymmetric matrix elements of $V$
we finally arrive at
\bea
&&
\delta_{v(x)} \varphi_k (x^\prime ) =
\sum_{l\atop \epsilon_l\ne \epsilon_k} \; 
\frac{1}{\epsilon_k-\epsilon_l}\; \varphi_l(x^\prime)
\; \sum_j^N \; \times
\label{invHFexpl}
\\
&&
\Bigg(\int\! dz\, \varphi^*_l(z)\varphi^*_j(z-x)
     \Big( \varphi_k(z)  \varphi_j(z-x)
          -\varphi_k(z-x)\varphi_j(z)  \Big)
\nonumber\\
&&
+\int\! dz\, \varphi^*_l(z)\varphi^*_j(z+x)
     \Big( \varphi_k(z)  \varphi_j(z+x)
          -\varphi_k(z+x)\varphi_j(z)  \Big)
\nonumber\\
&&
+\int\!dz\,dz^\prime\,
\varphi^*_l(z) \Big(\delta_{v(x)}\varphi^*_j(z^\prime)\Big)v(|z-z^\prime|)
     \Big( \varphi_k(z) \varphi_j(z^\prime)
          -\varphi_k(z^\prime)\varphi_j(z)\Big)
\nonumber\\
&&
+\int\!dz\,dz^\prime\,
\Big(\varphi^*_l(z) \varphi^*_j(z^\prime)
-\varphi^*_l(z^\prime) \varphi^*_j(z)\Big) v(|z-z^\prime|)
\varphi_k(z) \delta_{v(x)}\varphi_j(z^\prime)
\Bigg)
\nonumber
.
\eea
This linear equation can be solved directly
(where for Hamiltonian with real matrix elements in coordinate space
the orbitals, and thus their functional derivatives, 
can be chosen real)
or, quite effectively, by iteration, starting for example with initial guess
$\delta_{v(x)}\varphi_j(z^\prime)$ = 0.
As the $\delta_{v(x)}\varphi_k(x^\prime)$,
which are only required for the $N$ lowest orbitals,
depend on two position variables $x$, $x^\prime$,
Eq.\ (\ref{invHFexpl}) has essentially the dimension of a two--body equation.
Having calculated 
$\delta_{v(x)} \varphi_k (x_{i,l})$
=
$\Delta_{kl;x,i}$
(for $x > 0$, $1\le k\le N$, $1\le l\le N$, $1\le i\le n$)
from Eq.\ (\ref{invHFexpl})
the likelihood terms in the stationarity equation
(\ref{posterior-statEQ})
follow as
\bea
\delta_{v(x)} \ln p(x_i|\hat x,\rho_{\rm HF}(v))
&=&
\frac{\sum_{kl}^N M_{kl;i}\Delta_{kl;i}(x)}{\det B_i}
+
\frac{\sum_{kl}^N M^\dagger_{kl;i}
\Delta^\dagger_{kl;i}(x)}{\det B^\dagger_i} 
\nonumber\\&=&
 {\rm Tr} (B_i^{-1}\Delta_{i}(x))
+{\rm Tr} ({B_i^\dagger}^{-1}\Delta_{i}^\dagger(x))
,
\eea
recalling  that $M_{kl,i}$ = $(B_i)^{-1}_{lk}\det B_i$
and defining 
analogously to $B_i$
the matrix $\Delta_{i}(x)$ 
with elements $\Delta_{kl;i}(x)$.
The freedom to linearly
rearrange orbitals within
the Slater determinants 
$\det \{\ket{x_{i,l}}\}$
=
$\det \{\ket{\tilde x_{i,l}}\}$
(for each data point $i$, analogously for $\det \{\ket{\varphi_k}\}$),
makes it possible to diagonalize
the matrix of overlaps
$\scp{\varphi_k}{\tilde x_{i,l}}$
in new orbitals $\ket{\tilde x_{i,l}}$,
which are then linear combinations of the $\ket{x_{i,l}}$
\cite{Lemm-1994b,Lemm-1995a}.

\subsection{Numerical example of an inverse Hartree--Fock calculation}

\begin{figure}
\begin{center}
\psfig{file=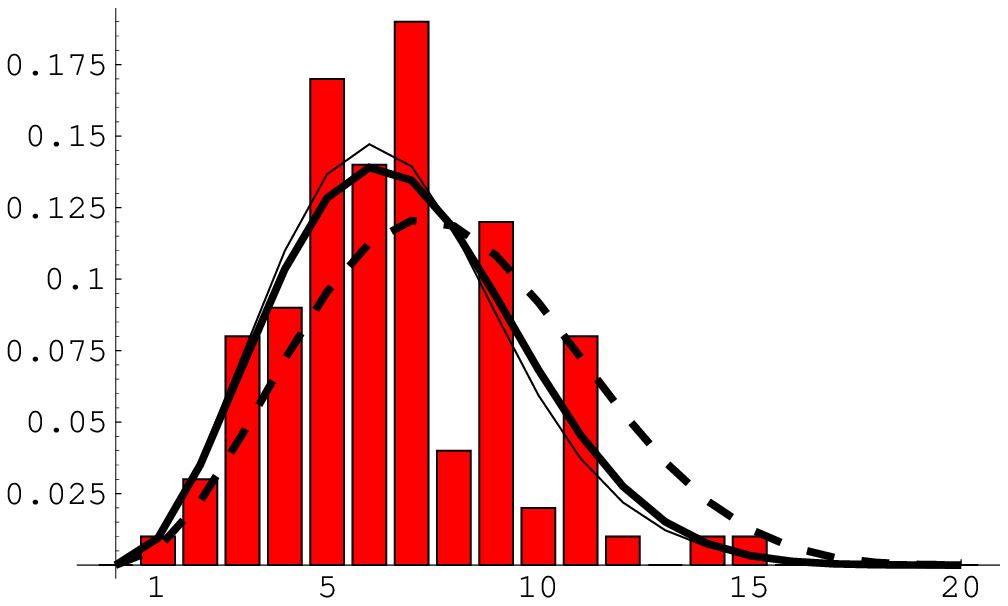, width= 67mm}
$\!\!\!$
\epsfig{file=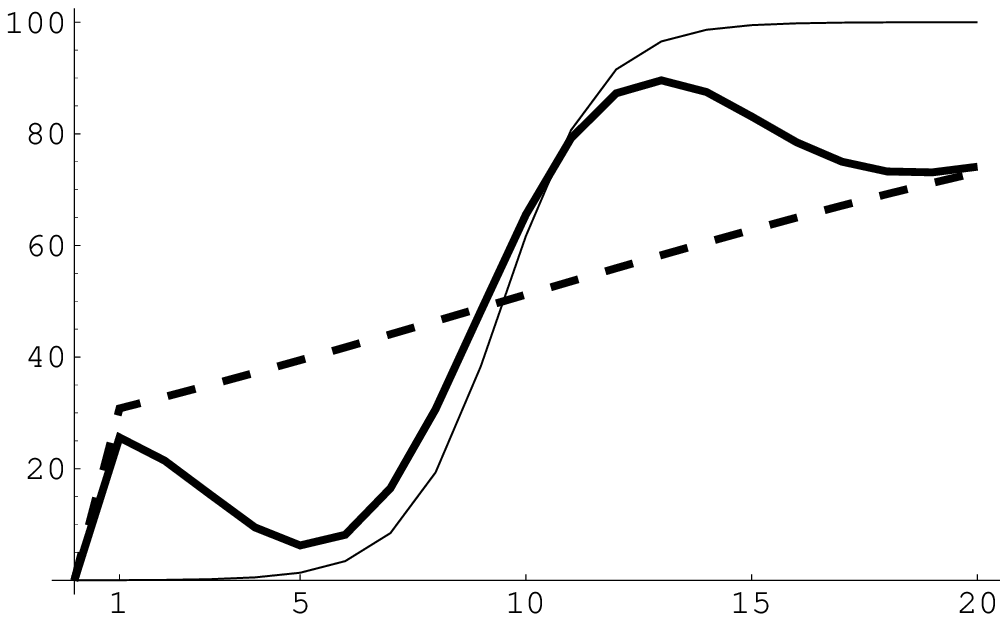, width= 67mm}
\setlength{\unitlength}{1mm}
\begin{picture}(0,0)
\put(-30,48){\makebox(0,0){relative likelihoods}}
\put(36.5,48){\makebox(0,0){two--body potentials}}
\put(34,12){\makebox(0,0){$v_{\rm true}$}}
\put(17,15){\makebox(0,0){$v$}}
\put(20,25){\makebox(0,0){$v_0$}}
\put(-30,0){\makebox(0,0){inter--particle distance}}
\put(36.5,0){\makebox(0,0){inter--particle distance}}
\end{picture}
\end{center}
\caption{Inverse Hartree--Fock Approximation:
The exact two--body likelihood has been calculated
for two one--dimensional particles
with many--body Hamiltonian $H$, given in Eq.\ (\ref{ex-two-body-H}),
a local one--body potential 
$V_1(x,x^\prime )$ = $\delta (x-x^\prime ) a (x/10)^2$, $a=10^{-3}$
breaking translational symmetry,
mass $m$ = $10^{-3}$,
and a given local two--body potential $V_{\rm true}$
of form (\ref{asymlocpot})
with 
$v_{\rm true}(|x-x^\prime|)$ = $b/(1+e^{-2 \gamma (x-k/2)/k})$,
$b$ = $100$, $\gamma$ = 10, $k$ = 21,
(thin line on the r.h.s.).
As training data
100 pairs $\{(x_{i,1},x_{i,2})|1\le i\le 100\}$
have been sampled according to that exact likelihood.
The corresponding exact (thin line) and empirical (bars) likelihoods
for inter--particle distances $|x_{i,1}-x_{i,2}|$
are shown on the  l.h.s.\ of the figure.
To reconstruct the potential 
a Gaussian prior has been used 
with $\lambda{\bf K}_0$ = $\lambda ({\bf I}-\Delta)/2$,
$\lambda$ = $0.5$ $10^{-3}$, 
and reference potential (dashed on r.h.s)
$v_{0}(|x-x^\prime|)$ = $b/(1+e^{-2 \gamma (x-k/2)/k})$,
$b$ = $100$, $\gamma$ = 1, $v_0(0)$ = 0.
The related reference likelihood of inter--particle distances 
is shown on the l.h.s.(dashed).
The reconstructed potential $v$
has been obtained by iterating with 
${\bf A}$ = ${\bf K}_0$ according to Eq.\ (\ref{iter2}) 
and solving Eqs.\ (\ref{HF-eq}) and (\ref{invHFexpl})
within each iteration step.
The problem has been studied at zero temperature,
$v$ fulfilling the boundary conditions 
$v(0)$ = 0 
and $v$ = constant beyond the right boundary.
No energy penalty term $E_U$ had to be included.
Note, that the number of data is not only small
for large inter--particle distances where the potential is large, 
but also for small distances.
This effect is due to antisymmetry which does not allow particles
to be at the same place.
Thus, the reconstructed potential $v$ is nearly equal 
to the reference potential $v_0$
for large and for small distances.
}
\label{hf-fig1}
\end{figure}

To test the numerical implementation of 
an inverse Hartree--Fock approach,
we study a two--body problem,
defined by the Hamiltonian
\be
H = -\frac{1}{2m}\Delta+V_1+V_{\rm true}
.
\label{ex-two-body-H}
\ee
Herein 
we assume the local one--body potential 
$V_1(x,x^\prime )$ = $\delta (x-x^\prime ) v_1(x)$
to be given and the two--body potential $V$
to be unknown, 
but local as in Eq.\ (\ref{asymlocpot}).
Hence, our aim is to approximate the function
$v(|x-x^\prime|)$, defining the matrix elements of $V$, 
by using empirical data
in combination with appropriate {\it a priori} information.

Fig.\ \ref{hf-fig1} 
shows the results of a corresponding
inverse Hartree--Fock calculation.
(The prior process
 and parameters are given in the figure caption, 
 computational details will be presented elsewhere).
For this two--body problem it is possible to calculate
the exact solution and corresponding likelihood numerically.
Hence, we was able to sample training data using the exact likelihood.
Note that, besides the problem of simulating realistic data,
an inverse Hartree--Fock calculation 
for more than two particles is not much more complex than for two particles.
It only requires to add one single particle orbital
for every additional particle. 
Thus, an analogous inverse Hartree--Fock calculation is clearly
computationally feasible for 
many--body systems with three or more particles.

We have already discussed in previous sections
that, in regions where the potential is large,  
the reconstruction of a potential is essentially based
on {\it a priori} information.
Training data are less important in such regions,
because finding a particle there is very unlikely.
In Fig.\ \ref{hf-fig1}, for example, 
{\it a priori} information is thus especially important for large distances.
A new, similar phenomenon occurs now when dealing with fermions:
The antisymmetry, we have to require for fermions,
forbids different particles to be at the same location.
Hence, antisymmetry 
reduces the number of training data for small distances,
and {\it a priori} information 
becomes especially important.
This effect can clearly be seen in the figure,
where the reconstructed potential $v$
is influenced by the data mainly for medium distances.
For large,
but also for small inter--particle distances,
the reconstructed potential
is quite similar to the reference potential.

Summarizing, we note that for inverse Hartree--Fock problems
in addition to the direct Hartree--Fock Eq.\ (\ref{HF-eq})
a second equation (\ref{invHFeq}) has to be solved
determining the change of Hartree--Fock orbitals under
a change of the potential.
Despite this complication it was possible to solve
the inverse Hartree--Fock equations numerically
for the example problem considered in this section.

\section{Conclusions}
\label{conclusions}

We have studied the inverse problem 
of reconstructing a quantum mechanical potential from empirical measurements.
The approach presented in this paper is based on Bayesian statistics
which has already been applied successfully 
to many empirical learning problems.
For quantum mechanical systems, 
empirical data enter the formalism through the likelihood function 
as defined by the axioms of quantum mechanics.
Additional {\it a priori} information 
is implemented in form of stochastic processes.
The reconstructed potential is then found 
by maximizing the Bayesian posterior density.

The specific advantage of this new, nonparametric Bayesian approach to 
inverse quantum theory
is the possibility 
to combine heterogeneous data,
resulting from arbitrary quantum mechanical measurements,
with a flexible and explicit implementation of {\it a priori} information.

Two numerical examples --- 
the reconstruction of an approximately periodic potential
and of a strictly symmetric potential --- have demonstrated
the computational feasibility of the Bayesian approach
for one--dimensional systems.
While a direct numerical solution is thus
possible for one--dimensional problems,
it becomes computationally demanding for two-- or three dimensional problems.

As a possible approximation scheme for many--body systems
an inverse Hartree--Fock approach has been proposed.
An implementation of a corresponding reconstruction algorithm
has been tested for a system of fermions,
for which we were able to solve the inverse Hartree--Fock equation numerically.

Finally, we want to emphasize
the flexibility of the Bayesian approach 
which can be easily adapted 
to a variety of different empirical learning situations.
This includes, as we have seen,
inverse problems in quantum theory
at zero and at finite temperature,
for single particles as well as for few-- or many--body systems.

\section*{Acknowledgements}

We are very grateful to A. Weiguny for many stimulating discussions.

\end{document}